\documentclass[12pt,twocolumn]{aastex63}
\usepackage{amsmath}

\newcommand{\kms}{\rm km~s^{-1}}
\newcommand{\kmsmpc}{\rm km~s^{-1}~Mpc^{-1}}
\newcommand{\dn}{D_{n}4000}

\usepackage{color}
\usepackage{multirow}
\usepackage{amsmath}

\begin{document}

\title{The HectoMAP Cluster Survey: Spectroscopically Identified Clusters and their Brightest Cluster Galaxies (BCGs)}

\author{Jubee Sohn}
\affiliation{Smithsonian Astrophysical Observatory, 60 Garden Street, Cambridge, MA 02138, USA}

\author{Margaret J. Geller}
\affiliation{Smithsonian Astrophysical Observatory, 60 Garden Street, Cambridge, MA 02138, USA}

\author{Ho Seong Hwang}
\affiliation{Astronomy Program, Department of Physics and Astronomy, Seoul National University, 1 Gwanak-ro, Gwanak-gu, Seoul 08826, Korea}

\author{Antonaldo Diaferio}
\affiliation{Universit{\`a} di Torino, Dipartimento di Fisica, Torino, Italy}
\affiliation{Istituto Nazionale di Fisica Nucleare (INFN), Sezione di Torino, via P. Giuria 1, I-10125 Torino, Italy}

\author{Kenneth J. Rines}
\affiliation{Department of Physics and Astronomy, Western Washington University, Bellingham, WA 98225, USA}

\author{Yousuke Utsumi}
\affiliation{Kavli Institute for Particle Astrophysics and Cosmology, SLAC National Accelerator Laboratory, Stanford University, 2575 Sand Hill Road, Menlo Park, CA 94025, USA}

\email{jubee.sohn@cfa.harvard.edu}

\begin{abstract}
We apply a friends-of-friends (FoF) algorithm to identify galaxy clusters and we use the catalog to explore the evolutionary synergy between BCGs and their host clusters. We base the cluster catalog on the dense HectoMAP redshift survey (2000 redshifts deg$^{-2}$). The HectoMAP FoF catalog includes 346 clusters with 10 or more spectroscopic members. We list these clusters and their members (5992 galaxies with a spectroscopic redshift). We also include central velocity dispersions ($\sigma_{*, BCG}$) for all of the FoF cluster BCGs, a distinctive feature of the HectoMAP FoF catalog. HectoMAP clusters with higher galaxy number density (80 systems) are all genuine clusters with a strong concentration and a prominent BCG in Subaru/Hyper Suprime-Cam images. The phase-space diagrams show the expected elongation along the line-of-sight. Lower-density systems include some false positives. We establish a connection between BCGs and their host clusters by demonstrating that $\sigma_{*,BCG}/\sigma_{cl}$ decreases as a function of cluster velocity dispersion ($\sigma_{cl}$), in contrast, numerical simulations predict a constant $\sigma_{*, BCG}/\sigma_{cl}$. Sets of clusters at two different redshifts show that BCG evolution in massive systems is slow over the redshift range $z < 0.4$. The data strongly suggest that minor mergers may play an important role in BCG evolution in these clusters ($\sigma_{cl} \gtrsim 300~\kms$). For systems of lower mass ($\sigma_{cl} < 300~\kms$), the data indicate that major mergers may play a significant role. The coordinated evolution of BCGs and their host clusters provides an interesting test of simulations in high density regions of the universe.
\end{abstract}

\section{Introduction}

Galaxy clusters, the most massive gravitational bound systems, continually accrete material from their surroundings. In fact, a substantial fraction of their accretion is recent; clusters increase their mass by a factor of two between $z \sim 0.5$ and the present (e.g., \citealp{Zhao09, Fakhouri10, vandenBosch14, Haines18}). Thus a dense redshift survey that explores this epoch can place interesting constraints on the co-evolution of clusters and their members (e.g., \citealp{Dressler84, Blanton09, Peng10, Haines13, Wetzel14, Gullieuszik15, Sohn20}).

Catalogs of clusters are the necessary foundation for studying clusters and their members. Several techniques yield galaxy cluster catalogs. For example, X-ray observations reveal large samples of galaxy clusters by tracing the X-ray emitting hot intracluster medium (e.g., \citealp{Edge90, Gioia90, Ebeling98, Ebeling10, Bohringer00, Bohringer01, Bohringer17, Pacaud16}). The intergalactic medium in rich clusters also distorts the cosmic microwave background spectrum (the Sunyaev-Zel'dovich (SZ) effect) providing another route to cluster identification (e.g., \citealp{Melin06, Vanderlinde10, Marriage11, Bleem15, PlanckCollaboration15, PlanckCollaboration16}). X-ray and the SZ observations of clusters not only detect clusters, but also provide a measure of the cluster mass. 

Identifying galaxy over-densities in optical and infrared (IR) imaging is a long-standing technique for obtaining large samples of clusters. Since the first systematic survey of cluster by \citet{Abell58}, many surveys have identified galaxy clusters photometrically based on various optical and infrared imaging surveys (e.g., \citealp{Zwicky68, Abell89, Gladders00, Koester07, Wen09, Hao10, Rykoff14, Oguri18, Gonzalez19}). 

Dense spectroscopic surveys enable a robust identification of cluster members. Redshift measurements of the individual galaxies in the cluster field clearly separate the cluster members and interlopers. Previous studies compile spectroscopic redshift measurements of galaxies in clusters identified by other methods (e.g., X-ray, optical, and IR imaging) to refine these cluster catalogs (e.g., \citealp{Rozo15, Clerc16, Sohn18b, Sohn18a, Rines18, Myles20, Kirkpatrick21}). Other studies identify galaxy over-densities or, equivalently, clusters in redshift space (e.g., \citealp{Huchra82, Eke04, Berlind06, Robotham11, Tago10, Tempel14}). These catalogs generally provide an estimate of the cluster velocity dispersion, a mass proxy that complements other estimates.

HectoMAP \citep{Geller15, Hwang16, Sohn21} is a large-scale redshift survey designed to study galaxy cluster evolution in the intermediate redshift where clusters grow by a factor of 2. HectoMAP covers $\sim55$ deg$^{2}$ of the sky with $\sim 2000$ redshifts deg$^{-2}$. This high density survey enables robust identification of galaxy clusters based only on spectroscopy. Here, we apply a friends-of-friends (FoF) algorithm to identify galaxy clusters in HectoMAP purely based on the spectroscopy. The resultant cluster catalog includes 346 systems with more than 5992 members with $z \leq 0.6$. 

The HectoMAP region is included in the Subaru/Hyper Suprime-Cam (HSC) Strategic Survey Program (SSP) project \citep{Miyazaki12, Aihara18}. The exquisite imaging combined with the dense spectroscopy provides a platform for exploring the co-evolution of the 346 FoF clusters and their BCGs. In addition to the redshifts, the HectoMAP survey provides central velocity dispersions for all of the BCGs in the FoF catalog. The redshift coverage and the mass range of the HectoMAP FoF clusters enable a clean exploration of the relationship between the cluster velocity dispersion and the central velocity dispersion of the BCG as a function of cluster velocity dispersion and redshift (e.g., \citealp{Sohn20}). This relationship is a  test of current simulations of the growth of structure in $\Lambda$CDM. 

We first introduce the HectoMAP redshift survey in Section \ref{sec:data}. We describe the cluster identification algorithm in Section \ref{sec:identification}. In Section \ref{sec:cat}, we introduce the HectoMAP cluster catalog, and we also explore the physical properties of the HectoMAP clusters. We then investigate the connection between HectoMAP clusters and their BCGs as a test of simulations (Section \ref{sec:connection} and Section \ref{sec:discussion}). We conclude in Section \ref{sec:conclusion}. We use the standard $\Lambda$CDM cosmology with $H_{0} = 70~\kmsmpc$, $\Omega_{m} = 0.3$, $\Omega_{\Lambda} = 0.7$, and $\Omega_{k} = 0.0$ throughout.

\section{HectoMAP}\label{sec:data}

HectoMAP is a dense redshift survey of the intermediate-age universe with a median redshift $z \sim 0.31$ \citep{Geller11,Geller15,Hwang16,Sohn21}. The survey field is located at $200 < $ R.A. (deg) $< 250$ and $42.5 < $ Decl. (deg) $< 44.0$, covering 54.64 deg$^{2}$ of the sky. The full survey includes $\sim 110,000$ spectroscopic redshifts and the typical galaxy number density is $\sim 2000$ deg$^{-2}$. 

HectoMAP is included in the Subaru/HSC SSP fields \citep{Miyazaki12, Aihara18}. \citet{Sohn21} published the spectroscopic data within 8.7 deg$^{2}$  that includes the HSC/SSP Data Release (DR) 1 coverage. \citet{Sohn21} described the details of the HectoMAP survey. Here we briefly review the photometric and spectroscopic data. 

\subsection{Photometry}\label{sec:phot}

SDSS DR16 \citep{SDSSDR16} is the photometric basis of HectoMAP. We select galaxies with SDSS $probPSF = 0$, where $probPSF$ indicates the probability that the object is a star. Following \citet{Sohn21}, we use Petrosian magnitudes for the galaxies and we compute galaxy colors based on model magnitudes. 

Because the HectoMAP survey covers a wide redshift range, we apply the $K-$correction to the galaxy photometry. We use the $kcorrect$ code \citep{Blanton07} to derive the $K-$correction at $z = 0.35$, the median redshift of the HectoMAP survey. Hereafter, we use galaxy magnitudes and colors after both foreground-extinction and K-correction. 

\subsection{Spectroscopy}\label{sec:spec}

The HectoMAP spectroscopy comes from two major spectroscopic surveys: SDSS/BOSS and our own MMT/Hectospec survey. We first compiled the SDSS DR16 spectroscopy which includes 25524 SDSS and BOSS redshifts within the HectoMAP field. The typical redshift uncertainty of these SDSS/BOSS measurements is $\sim 36~\kms$.

The majority of HectoMAP spectroscopy is from the Multi-Mirror Telescope (MMT)/Hectospec survey. Hectospec is a multi-object fiber-fed spectrograph mounted on the MMT 6.5m telescope \citep{Fabricant98, Fabricant05}. Hectospec has 300 fibers deployable over a 1 degree diameter field. A Hectospec spectrum, obtained through a 270 mm$^{-1}$ grating, covers the wavelength range 3700 - 9100 {\rm \AA~} with an average resolution of 6.2 {\rm \AA}. The Hectospec survey was carried out from 2009 to 2019. The primary targets of HectoMAP are galaxies with $r < 20.5$ and $(g-r) > 1$, and galaxies with $20.5 \leq r < 21.3$, $g-r > 1$, and $r-i > 0.5$. 

We reduce the Hectospec spectra using the standard HSRED v2.0 package\footnote{http://mmto.org/$\sim$rcool/hsred/}. We measured the redshift using cross-correlation (RVSAO, \citealp{Kurtz98}). We also visually inspected the cross-correlation results and classified them into three categories: `Q' for high quality fits, `?' for ambiguous fits, and `X' for poor fits. We use only redshifts with `Q' for further analysis. We note that the typical offset between the Hectospec and SDSS/BOSS redshifts is $\sim 26~\kms$ \citep{Sohn21}, less than the typical uncertainty in the Hectospec redshift ($\sim 40~\kms$)

\begin{figure}
\centering
\includegraphics[scale=0.47]{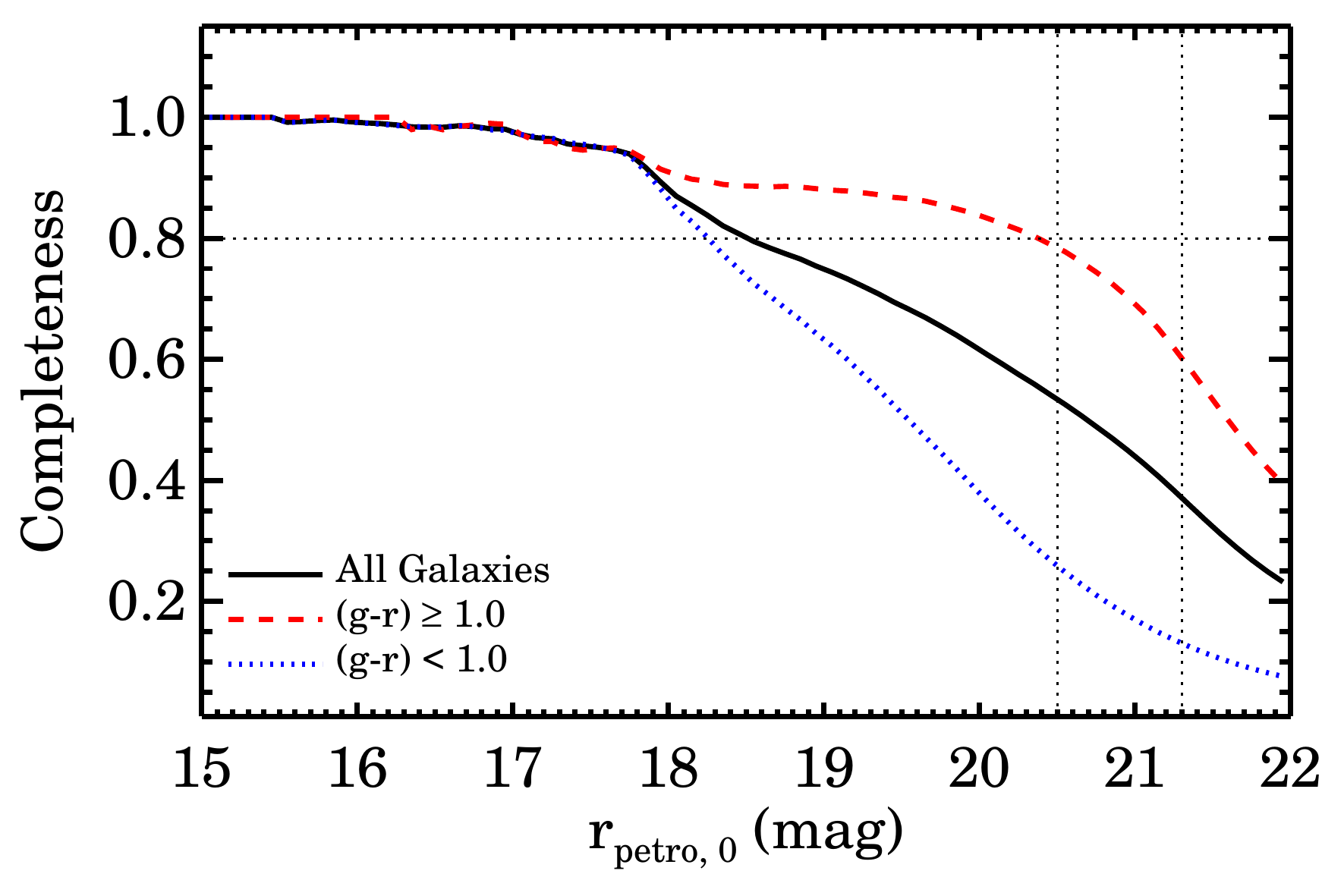}
\caption{Spectroscopic survey completeness of HectoMAP as a function of $r-$band magnitude. Black, red, and blue lines show the integral completeness of the entire, red ($g-r \geq 1.0$) and blue ($g-r < 1.0$) subsamples, respectively. }
\label{complete}
\end{figure}

Figure \ref{complete} shows the HectoMAP spectroscopic survey completeness as a function of $r-$band magnitude. The survey integral completeness for the main targets with $(g-r) > 1.0$ is 80\% at $r = 20.5$ and 62\% at $r = 21.3$. The survey is much less complete for bluer objects outside the target range. 

% ========================================================
% Figure \ref{complete}
% ========================================================
% ========================================================

We derive two additional spectroscopic properties of the HectoMAP galaxies: $\dn$ and the central stellar velocity dispersion. We first measure the $\dn$ index, a stellar population age indicator (e.g., \citealp{Kauffmann03}). Following the definition from \citet{Balogh99}, we compute the flux ratio between $4000 - 4100$ \AA~ and $3850 - 3950$ \AA: $\dn = F_{\lambda} (4000 - 4100) / F_{\lambda}(3850 - 3950)$. We use the $\dn$ index for characterizing brightest cluster galaxies in Section \ref{sec:connection}. 

We also derive the central stellar velocity dispersion of HectoMAP galaxies. For SDSS/BOSS spectra, we obtain the stellar velocity dispersion from the Portsmouth data reduction. The Portsmouth data reduction \citep{Thomas13} measures the velocity dispersion using the Penalized Pixel-Fitting (pPXF) code \citep{Cappellari04}. There are 10,992 HectoMAP galaxies with Portsmouth velocity dispersion measurements. 

For HectoMAP galaxies with MMT/Hectospec spectra, we estimate the velocity dispersion using the University of Lyon Spectroscopic analysis Software (ULySS, \citealp{Koleva09}). ULySS derives the velocity dispersion by comparing the observed spectra with stellar population templates based on the PEGASE-HR code and the MILES stellar library. We use the rest-frame spectral range $4100 - 5500$ \AA~ for deriving the stellar velocity dispersion to minimize the velocity dispersion uncertainty. A total of 91\% of quiescent galaxies in HectoMAP have a measured velocity dispersion.

Because the fiber sizes of Hectospec ($0.75\arcsec$ radius) and SDSS ($1.5\arcsec$ radius) differ, we apply an aperture correction. The aperture correction is defined as $\sigma_{A}/\sigma_{B} = (R_{A} / R_{B})^{\beta}$, where $\sigma$ is the stellar velocity dispersion, $R$ is the fiber aperture. We use the aperture correction coefficient $\beta = -0.054 \pm 0.005$ following \citet{Sohn17}. We correct the velocity dispersion to a fiducial physical aperture 3 kpc \citep{Zahid17, Sohn17, Sohn20}: $\sigma_{3 {\rm kpc}} / \sigma_{\rm SDSS/Hecto} = (3 {\rm kpc} / R_{SDSS/Hecto})^{\beta}$, where $R_{SDSS/Hecto}$ is the physical scale corresponding to SDSS/Hectospec aperture. We note that the median difference between the raw and aperture corrected velocity dispersions is small ($\sim 3\%$). In Section \ref{sec:connection}, we use these velocity dispersions to explore the relationship between the BCGs and their host clusters. Essentially all of the 346 BCGs have a measured velocity dispersion.

\section{Cluster Identification}\label{sec:identification}

Our first goal is to identify galaxy clusters and their members based on spectroscopy. We describe the friends-of-friends (FoF) algorithm we use for identifying galaxy systems (Section \ref{sec:fof}). We then elucidate the empirical determination of linking lengths for the FoF algorithm (Section \ref{sec:ll}). We describe the construction of the full HectoMAP FoF catalog in Section \ref{sec:iden}, and we explore the properties of the the cluster catalog in Section \ref{sec:explore}. 

\subsection{Friends-of-Friends Algorithm}\label{sec:fof}

The FoF algorithm \citep{Huchra82} has a long history as a tool for identifying clusters of galaxies. Starting from a galaxy, the algorithm finds neighboring galaxies (friends) within a given linking length and repeats this search for neighbors of the neighbors (friends of friends). The set of connected neighboring galaxies constitute a single galaxy system. 

The FoF algorithm is straightforward to apply to large surveys. Furthermore, the algorithm does not require any {\it a priori} physical assumptions about the galaxy systems including, but not limited to, their three dimensional geometry or their number density profile \citep{Duarte14}. Many previous studies build catalogs of galaxy systems using the FoF algorithm (e.g., \citealp{Huchra82, Barton96, Eke04, Berlind06, Tago10, Robotham11, Tempel12, Tempel14, Tempel16, Hwang16, Sohn16, Sohn18b}); these catalogs include galaxy systems on various scales from groups (e.g., \citealp{Ramella97, Sohn16}) to the large scale features in the cosmic web (e.g., \citealp{Hwang16}). 

We apply the FoF algorithm in redshift space. The standard FoF algorithm \citep{Huchra82} in redshift space requires two linking lengths: one in the projected spatial direction ($\Delta D$) and one in the radial direction ($\Delta V$). We connect two galaxies if the separation between them in both the projected spatial and radial directions are smaller than the relevant linking lengths. We define the linking lengths as: 
\begin{equation}
\Delta D = b_{proj} \times \bar{n}_{g} (z)^{-1/3}, 
\end{equation}
and
\begin{equation}
\Delta V = b_{radial} \times \bar{n}_{g} (z)^{-1/3},
\end{equation}
where $\bar{n}_{g} (z)$ is the mean galaxy volume number density of the survey (generally a function of redshift $z$), and $b_{proj}$ and $b_{radial}$ are the projected spatial  and radial linking lengths in units of the mean galaxy separation within the survey at redshift $z$. 

The choice of linking length determines the nature of galaxy systems that the FoF algorithm identifies. For example, if the linking length is too large, galaxies that are not physically connected can be bundled into a galaxy system. In contrast, the FoF algorithm with a tight linking length breaks galaxy systems into smaller fragments and thus the algorithm detects only dense, compact systems. Despite its importance, the determination of optimal linking lengths is not straightforward \citep{Duarte14}. 

The projected linking length determines the density contrast of systems identified by the FoF algorithm. \citet{Huchra82} demonstrate that the minimum galaxy overdensity of the FoF systems depends on the projected linking length: 
\begin{equation}
\frac{\delta n}{n} = \frac{3}{4\pi b_{proj}^{3}} - 1.     
\end{equation}
\citet{Duarte14} compare the FoF linking lengths used in various catalogs (see their Table 1). The minimum overdensity of previous FoF cluster surveys varies from 80 - 1100, corresponding to $0.06 < b_{proj} < 0.14$. A smaller projected linking length identifies denser systems. We use a projected linking length within this range; $b_{proj} \sim 0.13$ corresponds to a minimum overdensity of 110 (see Section \ref{sec:ll}). 

Many previous cluster surveys based on large redshift surveys and the FoF algorithm use variable linking lengths to cover the survey redshift range (e.g., \citealp{Huchra82, Eke04, Robotham11, Duarte14, Tempel16}). In general, the galaxy number density ($\bar{n}_{g} (z)$) varies as a function of redshift in a magnitude-limited redshift survey. Thus, the FoF algorithm with a fixed $b_{proj}$ and $b_{radial}$ identifies neighboring galaxies with different densities and density contrasts at different redshifts. Varying the linking length identifies systems with similar over-densities over the redshift survey range. One issue with this approach is that at the limiting redshift of the survey where the mean galaxy density drops, the FoF bundles large numbers of unrelated galaxies into single extended systems.

\begin{figure}
\centering
\includegraphics[scale=0.47]{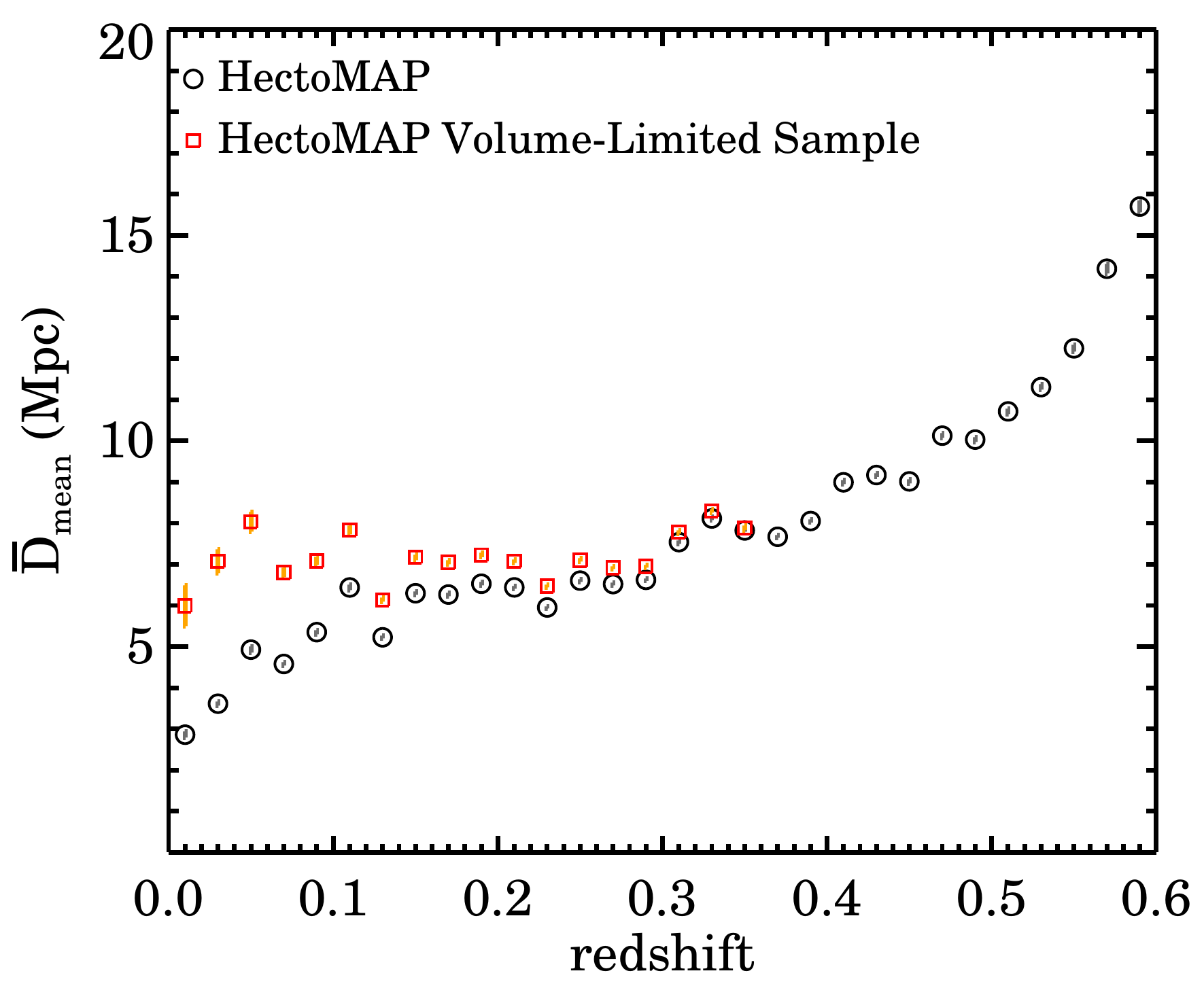}
\caption{Mean separation of all HectoMAP galaxies (black circles) and HectoMAP galaxies in the volume-limited sample (red squares) as a function of redshift. }
\label{dmean}
\end{figure}

Figure \ref{dmean} displays the mean separation ($\overline{D}_{mean} (z)$) of HectoMAP galaxies as a function of redshift. We compute  $\overline{D}_{mean} (z)$ from the mean number density ($\bar{n}_{g} (z)$) in each redshift bin: $\overline{D}_{mean} (z) = \bar{n}_{g} (z)^{-1/3}$. As a result of the survey selection that is not purely magnitude limited, the mean separation of HectoMAP galaxies remains constant over the redshift range for $0.1 \lesssim z \lesssim 0.45$ and increases only beyond $z > 0.45$. The HectoMAP survey density drops rapidly at $z > 0.45$. This decrease in the survey number density occurs when the $20.5 < r < 21.3$ sample dominates the survey. 

% ========================================================
% Figure \ref{dmean}
% ========================================================
% ========================================================

\begin{figure}
\centering
\includegraphics[scale=0.47]{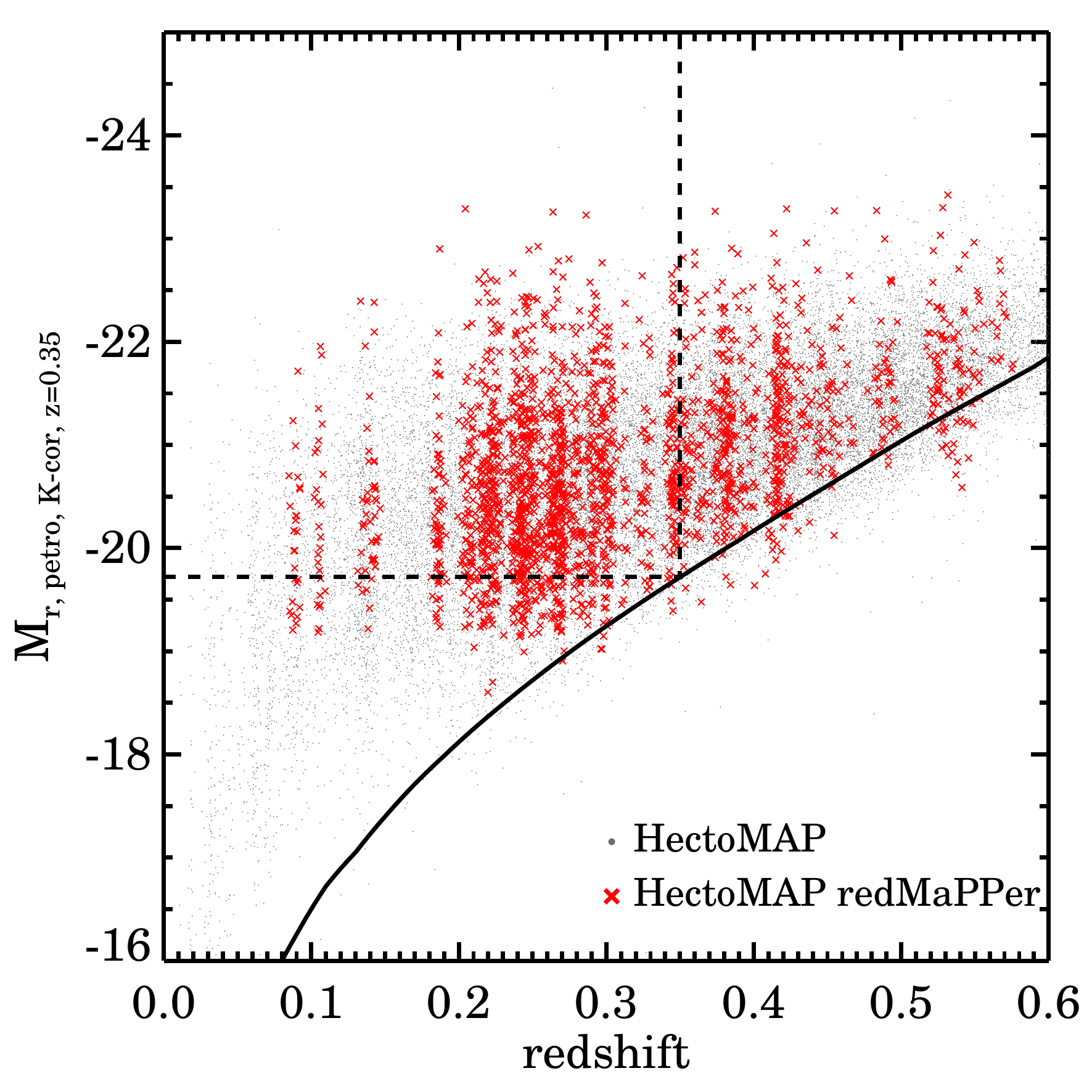}
\caption{Foreground extinction- and $K-$corrected $r-$band absolute magnitudes of HectoMAP galaxies (gray points) as a function of redshift. We plot only 25\% of galaxies for clarity. Red crosses indicate the spectroscopically identified redMaPPer members in HectoMAP. The solid line indicates the survey magnitude limit, $r = 21.3$. Dashed lines mark the boundary of the distance-limited sample ($M_{r} \leq -19.72$ and $z \leq 0.35$). }
\label{rmmagz}
\end{figure}

Applying the FoF algorithm to a volume-limited subsample is insensitive to selection biases introduced by the change in survey number density. Figure \ref{rmmagz} displays the foreground extinction- and K-corrected $r-$band magnitude of HectoMAP galaxies as a function of redshift. We derive the survey limit that corresponds to $r = 21.3$ based on the median foreground extinction- and K-correction as a function of redshift (the solid line). We then define a volume-limited sample with $z < 0.35$ and $M_{r} < -19.72$ (the dashed lines). In Figure \ref{dmean}, red squares show the mean survey density of galaxies in the volume-limited sample. Indeed, the mean separation is constant within the redshift range of the volume-limited sample. In particular, the mean number density of the volume-limited sample does not decrease at $z < 0.1$ unlike the mean density of the full sample. Thus the linking length in the volume-limited sample is constant throughout the survey redshift range. 

Because the volume limited sample covers most of the survey redshift range, we extend the FoF algorithm with a fixed linking length from the volume-limited sample throughout the survey (see Section \ref{sec:ll}). This approach enables identification of galaxy systems with similar physical properties \citep{Barton96, Sohn16}. Section \ref{sec:explore} discusses the systematics introduced by this choice.

% ========================================================
% Figure \ref{hecs_magz}
% ========================================================
% ========================================================

\subsection{Empirical Determination of the Linking Length}\label{sec:ll}

We use the redMaPPer clusters \citep{Rykoff14, Rykoff16} as a training set for empirical determination of linking length. redMaPPer (hereafter RM) is a photometric cluster finding algorithm based on the red-sequence. The RM catalog includes a large number of systems over a wide mass range and it is unbiased by selection of the BCG. The RM catalog (v6.3) based on the SDSS DR8 \citep{Rykoff16} lists 104 systems in the HectoMAP region. These HectoMAP RM systems are a sufficient basis for an empirical test of the success rate of the FoF algorithm as a function of the linking length. 

We previously tested the fidelity of the HectoMAP RM clusters based on our redshift survey \citep{Sohn18b, Sohn21}. Over $90\%$ of the HectoMAP RM clusters are genuine clusters with 10 or more spectroscopic members. The typical number of spectroscopically identified members of these RM systems is $\sim 20$ \citep{Sohn18b}. Thus, we can ask which set of linking lengths recovers these populous clusters.

\begin{figure}
\centering
\includegraphics[scale=0.47]{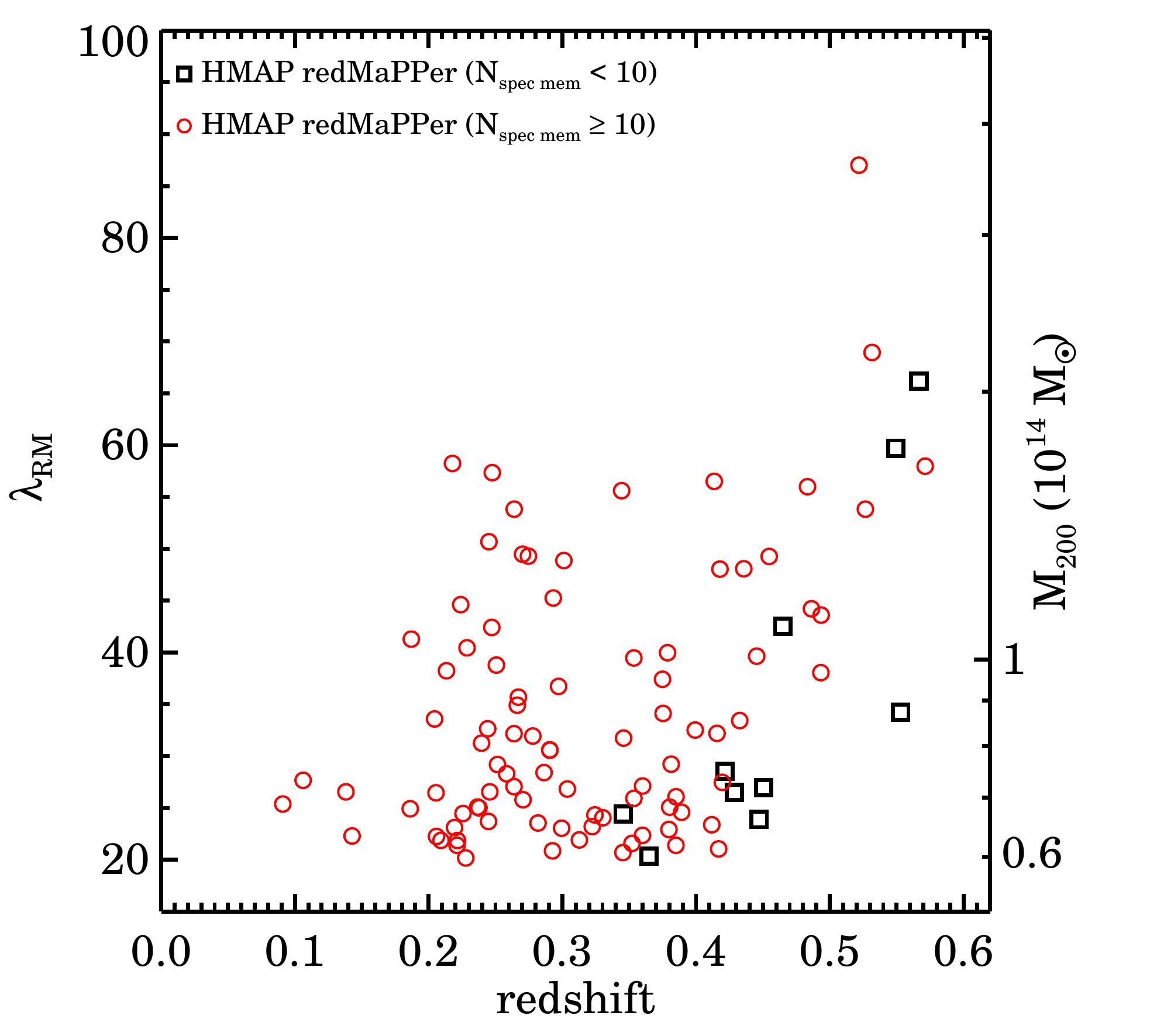}
\caption{Mass vs. redshift of redMaPPer clusters in HectoMAP with 10 or more members (red circles) and with less than 10 members (black squares). }
\label{rmz}
\end{figure}

The RM catalog also allows us to find the proper linking lengths for identifying low mass clusters. Figure \ref{rmz} shows the mass distribution ($M_{200}$, the mass enclosed within the radius where the density equals 200 times the critical density) of the HectoMAP RM clusters as a function of redshift. We compute $M_{200}$ using the relation between mass and RM richness \citep{Rines18}. The relation is based on 27 RM clusters with large richness ($\lambda > 64$) and with dense spectroscopy. In Figure \ref{rmz}, red circles show RM clusters with 10 or more spectroscopic members, and black squares indicate less populous systems. 

The HectoMAP RM sample includes clusters with M$_{200} \gtrsim 6 \times 10^{13}$ M$_{\odot}$ at $0.08 < z < 0.35$ where the redshift range corresponds to our volume-limited sample. An empirical test based on the HectoMAP RM clusters will find linking lengths that identify systems with mass larger than $6 \times 10^{13}$ M$_{\odot}$. The final sample we use for the empirical test includes 57 RM systems at $z < 0.35$ with 10 or more spectroscopic members. 

% ========================================================
% Figure \ref{hecsz}
% ========================================================
% ========================================================

For the empirical test, we generate a set of linking lengths by varying the projected linking lengths from 100 kpc to 1 Mpc in steps of 100 kpc. We explore radial linking lengths in the range $100~\kms$ to $1000~\kms$ in steps of $100~\kms$. We thus test 100 combinations of linking lengths to find the linking lengths that recovers the largest number of HectoMAP RM clusters. 

% ========================================================
% Figure \ref{hecs_magz}
% ========================================================
\begin{figure}
\centering
\includegraphics[scale=0.47]{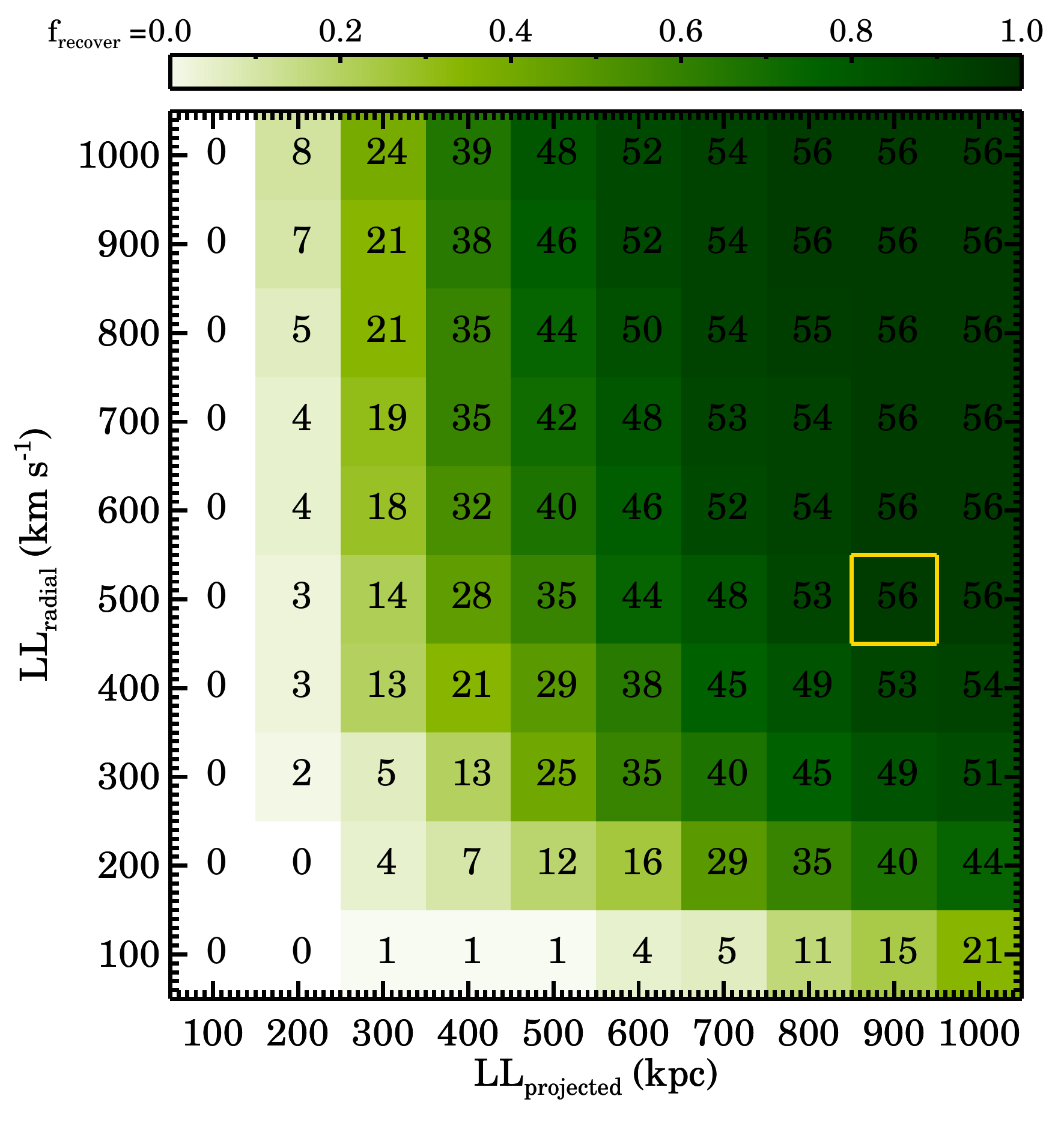}
\caption{Recovery rate of the HectoMAP redMaPPer systems by the FoF algorithm with various linking lengths. The x- and y-axes show the projected spatial and the radial linking lengths respectively. Darker colors indicate that more redMaPPer clusters are recovered.  }
\label{RM_recovery}
\end{figure}
% ========================================================

Figure \ref{RM_recovery} illustrates the result of the empirical test. The axes indicate the projected and radial linking lengths we test. In each pixel, we list the number of RM clusters recovered. With tighter linking lengths, the FoF algorithm misses many RM systems. The number of recovered RM systems also decreases slightly with the largest linking lengths (e.g., $\Delta D > 1000$ kpc or $\Delta V > 800~\kms$), because the algorithm bundles independent RM clusters into a single system. 

Based on the empirical test, we use linking lengths of 900 kpc and $500~\kms$ for identifying HectoMAP galaxy systems. We use the smallest radial linking length that recovers more than 90\% of the RM clusters. This catalog contains 248 systems with 10 or more spectroscopic members. These systems include all of the RM clusters except one with low galaxy number density; this missing RM cluster has an FoF counterpart with 6 members. The projected linking length corresponds to $b_{proj} \simeq 0.13$ (i.e., $\delta n / n \sim 110$), similar to linking lengths in a previous search for galaxy clusters based on 2dFGRS \citep{Eke04} or SDSS \citep{Berlind06}. 

The cluster identification based on a volume limited sample omits fainter cluster members. We remedy this drawback by selecting additional spectroscopic members within a cylindrical volume around the FoF cluster center (see Section \ref{sec:explore}). 

We also test the empirical linking lengths based on the HectoMAP X-ray clusters. \citet{Sohn18b} used ROSAT All-Sky survey data to identify 15 X-ray clusters in HectoMAP complete to limiting flux of $f_{X} = 3 \times 10^{-13}$ erg s$^{-1}$ cm$^{-2}$. All 15 X-ray clusters are successfully recovered by the choice of linking lengths. 

\begin{figure*}
\centering
\includegraphics[scale=0.60]{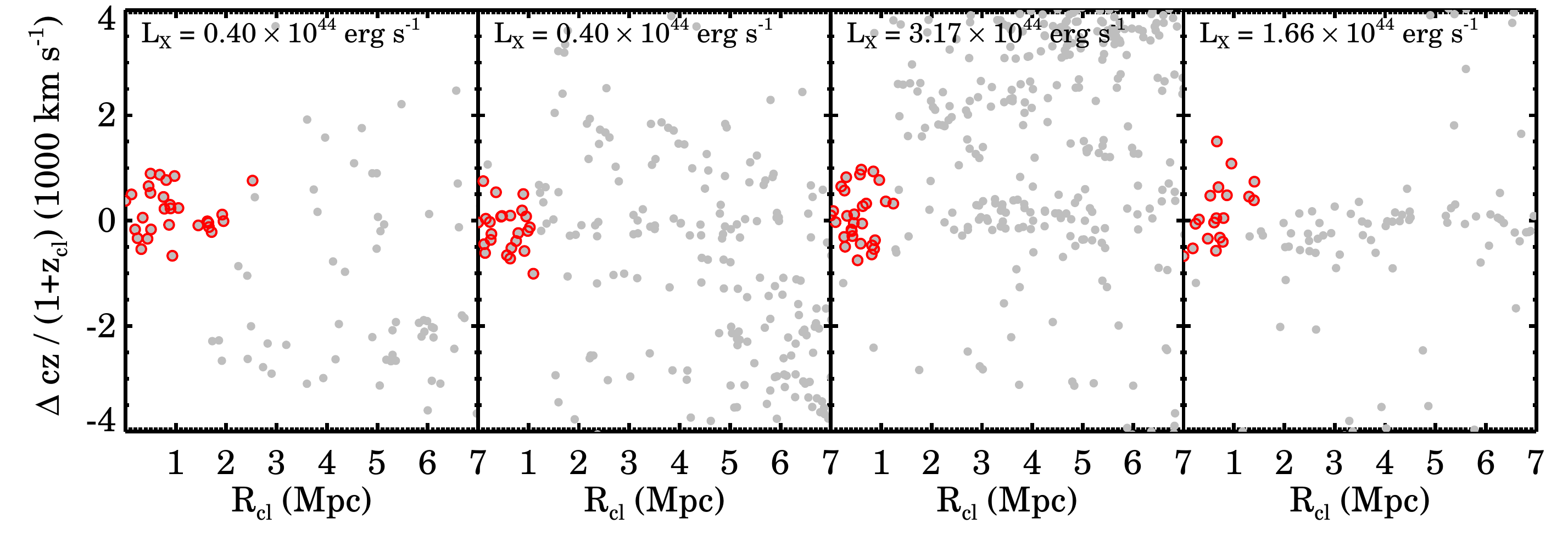}
\caption{Phase-space diagrams of the 4 HectoMAP X-ray clusters identified by the FoF algorithm, but not by the redMaPPer algorithm. Gray circles are the spectroscopic galaxies around the X-ray cluster center. Red circles are the members of the FoF clusters matched with the X-ray cluster. }
\label{xray_RM_missing}
\end{figure*}

Interestingly, five of the HectoMAP X-ray clusters are not included in the RM cluster catalog; one of them at $z = 0.03$ is out of the RM cluster survey redshift range. Figure \ref{xray_RM_missing} displays phase-space diagrams of the 4 X-ray clusters missing from RM. These phase-space diagrams, often referred to as the R-v diagram, show the relative rest-frame line-of-sight velocity difference versus the projected distances from the cluster center. In Figure \ref{xray_RM_missing}, gray and red circles show the spectroscopic galaxies around the X-ray cluster center and the FoF cluster members, respectively. The FoF algorithm identifies the spectroscopic members of the X-ray clusters successfully. 

% ========================================================
% Figure \ref{fof_summary}
% ========================================================
% ========================================================

Based on the Subaru/HSC SSP dataset, \citet{Jaelani20} identify a large sample of strong gravitational lens candidates including 13 candidates in HectoMAP that are within the magnitude range of the volume limited redshift survey. They visually identify strong lensing arcs around the center of photometrically identified CAMIRA clusters \citep{Oguri14, Oguri18}. We cross-match these 13 HectoMAP strong lensing candidates with the FoF cluster catalog. Ten of these strong lens candidates have HectoMAP FoF cluster counterparts. The other 3 systems have a FoF group counterpart with 4, 6, and 9 FoF members, respectively. These systems provide an additional test of the efficacy of the FoF cluster identification.

\subsection{Construction of the Full HectoMAP FoF Catalog}\label{sec:iden}

We extend the FoF catalog to cover the full redshift range of HectoMAP by applying the linking lengths determined from the volume-limited subset to higher redshifts $0.35 < z < 0.6$ where essentially all of the galaxies are intrinsically brighter than the limit for the volume limited sample of Section \ref{sec:ll}. The mean survey density remains constant for $0.35 < z < 0.45$ (Figure \ref{dmean}), slightly higher than the redshift limit of the volume-limited sample. At higher redshift $0.45 < z < 0.6$, the FoF clusters we identify with the fiducial linking lengths tend to be denser than their counterparts at lower redshift, an expected systematic. Increasing the linking lengths at the largest redshifts would lead to a large number of false positives because of the steep decline in the survey density. The FoF cluster catalog we construct for $z > 0.45$ still contains robust massive systems.

The FoF algorithm identifies a total of 12195 systems with more than two members in the full  HectoMAP survey. Most of these systems are pairs (59\%), triplets (20\%), or groups ($4 \leq N < 10$, 18\%). Following previous approaches \citep{Lee04, Sohn16, Sohn18b}, we further explore 346 systems with 10 or more FoF members (hereafter FoF clusters); 248 of these systems are within the volume limited subsample. The typical number of FoF members in these clusters is $\sim 17$. FoF systems with 10 or more members potentially contain many more faint members below the magnitude limit (see below).

We determine the center of each FoF cluster based on the center of light method \citep{Robotham11}. The center of light is basically identical to the center of mass, but it is based on galaxy luminosity rather than galaxy mass. We compute the center of light among FoF members. We then iterate after excluding the most distant FoF members from the center until only two members remain. Finally, we select the brighter galaxy as the system center; we define this central galaxy as the brightest cluster galaxies (BCGs) hearafter. For a majority ($\sim 75\%$) of the HectoMAP FoF clusters, the center corresponds to the location of the brightest cluster member. We discuss the properties of systems where the central galaxy is not the brightest cluster member in Section \ref{sec:connection}. Hereafter, we refer to the center of light as the cluster center. 

For the clusters identified in the volume-limited sample, there are members fainter than the magnitude limit ($M_{r} = -19.72$) that are not included by the FoF. We identify these faint members within $R_{cl} < max (R_{proj, FoF})$ and $|c (z_{galaxy} - z_{cl}) / (1 + z_{cl})| < max(|\Delta V_{FoF}|)$. Here, $max (R_{proj, FoF})$ is the largest projected distance of the FoF members, and $max (|\Delta V_{FoF}|)$ is the maximum radial velocity difference between the FoF members and the cluster center. We added $\sim 5$ faint members per cluster. We include these additional faint members in our analysis (e.g., to determine the cluster velocity dispersion).  

\subsection{Exploring the HectoMAP FoF Clusters}\label{sec:explore}

Taking advantage of the dense spectroscopy, we identify galaxy overdensities in redshift space as galaxy clusters. Like any method, the FoF algorithm does produce some false positives (e.g., \citealp{Ramella97, Diaferio99}). The algorithm may identify weak concentrations of galaxies or cuts through the extended filamentary structures where the central line-of-sight velocity dispersion is the value in the surrounding region. The inclusion of these features is unavoidable in constructing a cluster catalog based on the FoF algorithm. We thus explore the FoF cluster identification based on additional physical parameters. 

We use the galaxy number density to test the cluster identification because a high galaxy number density within the central region is a key characteristic of galaxy clusters. Additionally, we use deep Subaru/HSC imaging as a guide to the nature of the system. The HSC images also allow us to examine the morphology of the BCGs. The presence of extended quiescent early-type BCGs is characteristic of galaxy clusters.  

We compute the central galaxy number density ($\rho_{cl}$) for each cluster: $\rho_{VL} = N_{\rm galaxy} (M_{r} < -19.72) / V$. Here, $N_{\rm galaxy} (M_{r} < -19.72)$ is the number of galaxies brighter than $M_{r} = -19.72$ (this absolute magnitude limit is fainter than the survey limit for $z \gtrsim 0.35$), the magnitude limit of the volume-limited sample, $V$ is the cylindrical volume within $R_{proj} < 1$ Mpc, and $|c(z_{\rm galaxy} - z_{cl})| / (1 + z_{cl}) < 1000~\kms$. We compute the volume within $R_{proj} < 1$ Mpc, corresponding to the typical $R_{200}$ of galaxy clusters with $M_{200} > 10^{14} M_{\odot}$. The radial length of the cylindrical volume is also sufficient to encompass most spectroscopic members of the clusters. To compute the density contrast, we derive the galaxy number density of the entire HectoMAP survey also as a function of redshift. 

% ========================================================
% Figure \ref{fof_density}
% ========================================================
\begin{figure*}
\centering
\includegraphics[scale=0.65]{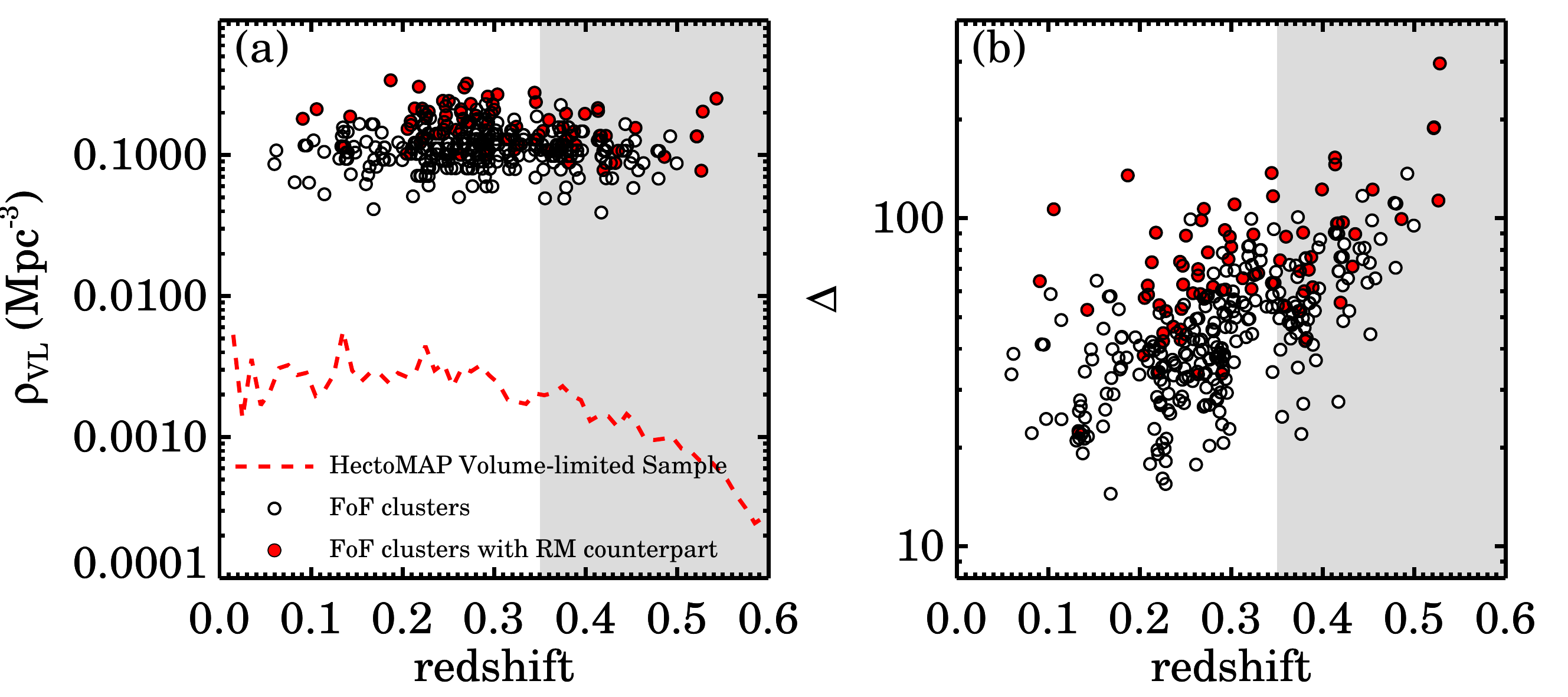}
\caption{(a) Galaxy number density of the FoF clusters (black circles) as a function of redshift. The red dashed line shows the mean number density of galaxies in the HectoMAP volume-limited sample. The gray shaded area indicates the redshift range where the survey limit is brighter than $M_{r} = -19.72$. (b) The density contrast of the FoF clusters as a function of redshift. }
\label{fof_density}
\end{figure*}
% ========================================================

Figure \ref{fof_density} (a) displays the central galaxy number density of the FoF clusters as a function of cluster redshift. The dashed line shows the average galaxy number density in the HectoMAP survey.  Figure \ref{fof_density} (b) shows the density contrast between the clusters and the HectoMAP survey density at the cluster redshift: $\Delta = \rho_{VL} / \rho_{HectoMAP} (z_{cl})$. Indeed, the FoF clusters have high density contrast ($\Delta > 10$) as expected. The density of the high-z ($z >0.35$) clusters generally exceeds the low-z ($z <0.35$) cluster densities because the FoF algorithm preferentially identifies higher density and higher density contrast clusters at high-z, where the survey density decreases.

In Figure \ref{fof_density}, red circles show FoF clusters with a RM counterpart. The RM clusters generally have higher number density although they are distributed over a wide density range. It is interesting that even at the highest galaxy number densities, there are FoF clusters (open circles) that are not identified by RM. We discuss these systems further below.

We compute the velocity dispersion of the FoF members as a cluster mass proxy. We use the bi-weight technique \citep{Beers90}, which yields a robust velocity dispersion measurement with a small number of members. The uncertainty in the velocity dispersion corresponds to the $1\sigma$ standard deviation derived from 1000 bootstrap resamplings. The typical uncertainty in the cluster velocity dispersion is $\sim 80~\kms$. 

\begin{figure}
\centering
\includegraphics[scale=0.43]{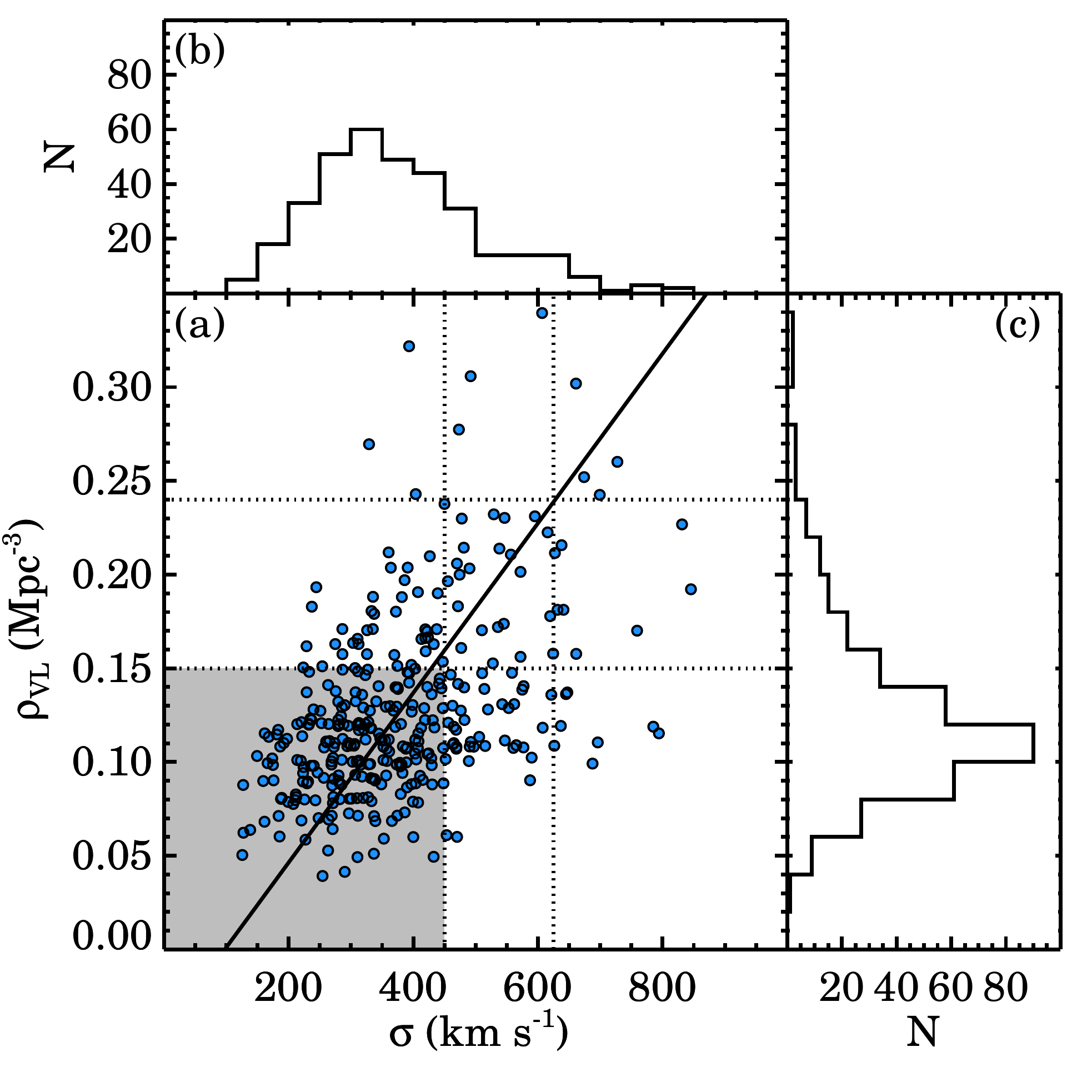}
\caption{(a) Galaxy number density within the FoF clusters as a function of the cluster velocity dispersion. The solid line shows the best-fit linear relation. The gray shaded area indicates the low density range which may include false positives. (b) The distribution of the cluster velocity dispersion and (c) the distribution of the galaxy number density.  }
\label{fof_density_sigma}
\end{figure}

Figure \ref{fof_density_sigma} (a) displays the galaxy number density as a function of the cluster velocity dispersion. Figure \ref{fof_density_sigma} (b) and (c) show the distributions of the number density and the cluster velocity dispersion, respectively. In general, the larger velocity dispersion (more massive) systems have higher galaxy number density. The Spearman's rank correlation coefficient is 0.45 with a significance of $1.13 \times 10^{-18}$. The solid line in Figure \ref{fof_density_sigma} shows the best-fit linear relation: $\rho_{VL} = (-0.044 \pm 0.014) + (0.091 \pm 0.008) \times (\sigma / 200 [\kms])$. According to this relation, a galaxy number density of 0.15 Mpc$^{-3}$ corresponds to a cluster velocity dispersion of $\sim 450~\kms$ and thus a cluster mass $\sim 10^{14} M_{\odot}$ \citep{Rines13}. This cluster mass is the approximate redMaPPer completeness limit and we thus use it for further exploration of the catalogs.

% ========================================================
% Figure \ref{fof_regime}
% ========================================================
% ========================================================

\begin{figure}
\centering
\includegraphics[scale=0.49]{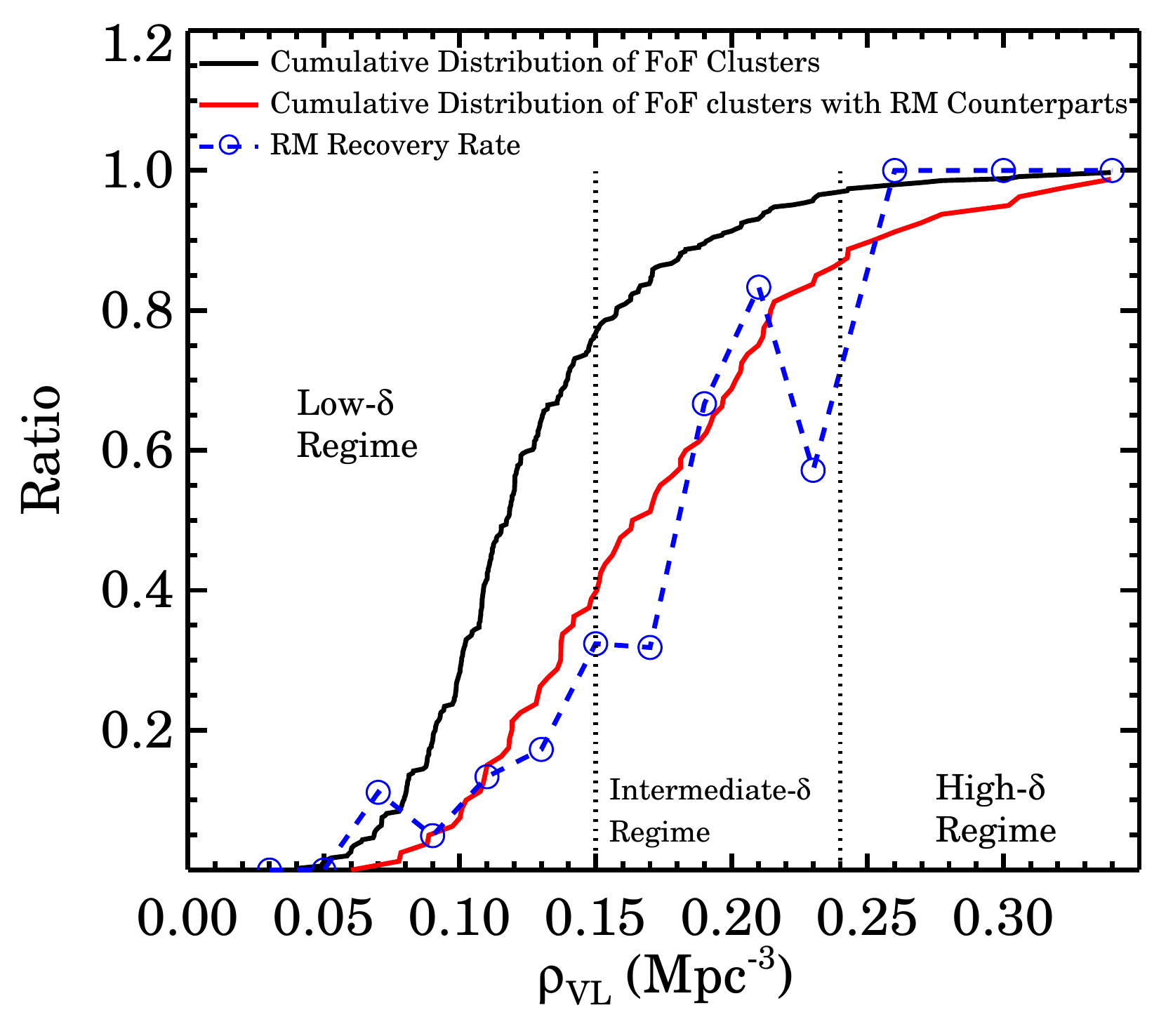}
\caption{Cumulative distribution of FoF clusters as a function of galaxy number density (black solid line). The red solid line shows the same distribution for FoF clusters with  redMaPPer counterparts. Blue circles and the blue dashed line shows the fraction of FoF clusters with redMaPPer counterparts binned in galaxy number density. In the high-density regime, every FoF cluster has a redMaPPer counterpart. In the intermediate-density regime, $\sim 55\%$ of FoF clusters have redMaPPer counterparts. Only $\sim10\%$ of FoF clusters have redMaPPer counterparts in the low-density regime. }
\label{fof_regime}
\end{figure}

Figure \ref{fof_regime} shows the cumulative distribution of the FoF cluster number density (the black line) ($\rho_{VL}$). The red solid line displays the cumulative distribution for the FoF clusters with a RM counterpart. We compute the fraction of FoF clusters with RM counterparts in three broad galaxy number density bins (the blue symbols in Figure \ref{fof_regime}). The three bins are high-density ($\rho_{VL} > 0.24$), intermediate-density ($0.15 < \rho_{VL} < 0.24$), and low-density ($\rho_{VL} < 0.15$). The clusters in the high-density regime generally have $\sigma \gtrsim 625~\kms$ ($M_{200} \gtrsim 2.5 \times 10^{14} M_{\odot}$) and those in the intermediate-density regime have $\sigma \gtrsim 450~\kms$ (Figure \ref{fof_density_sigma}). 

% ========================================================
% Figure \ref{fof_regime}
% ========================================================
% ========================================================

There are 10, 70, and 266 FoF clusters in the high-, intermediate-, and low-density regime, respectively. All of the clusters in the high-density regime have an RM counterpart supporting both approaches to cluster identification. In the intermediate density regime, 38 (54\%) FoF clusters have a RM counterpart. Within the intermediate density range the fraction of FoF clusters with RM counterparts increases with density. 

\begin{figure}
\centering
\includegraphics[scale=0.40]{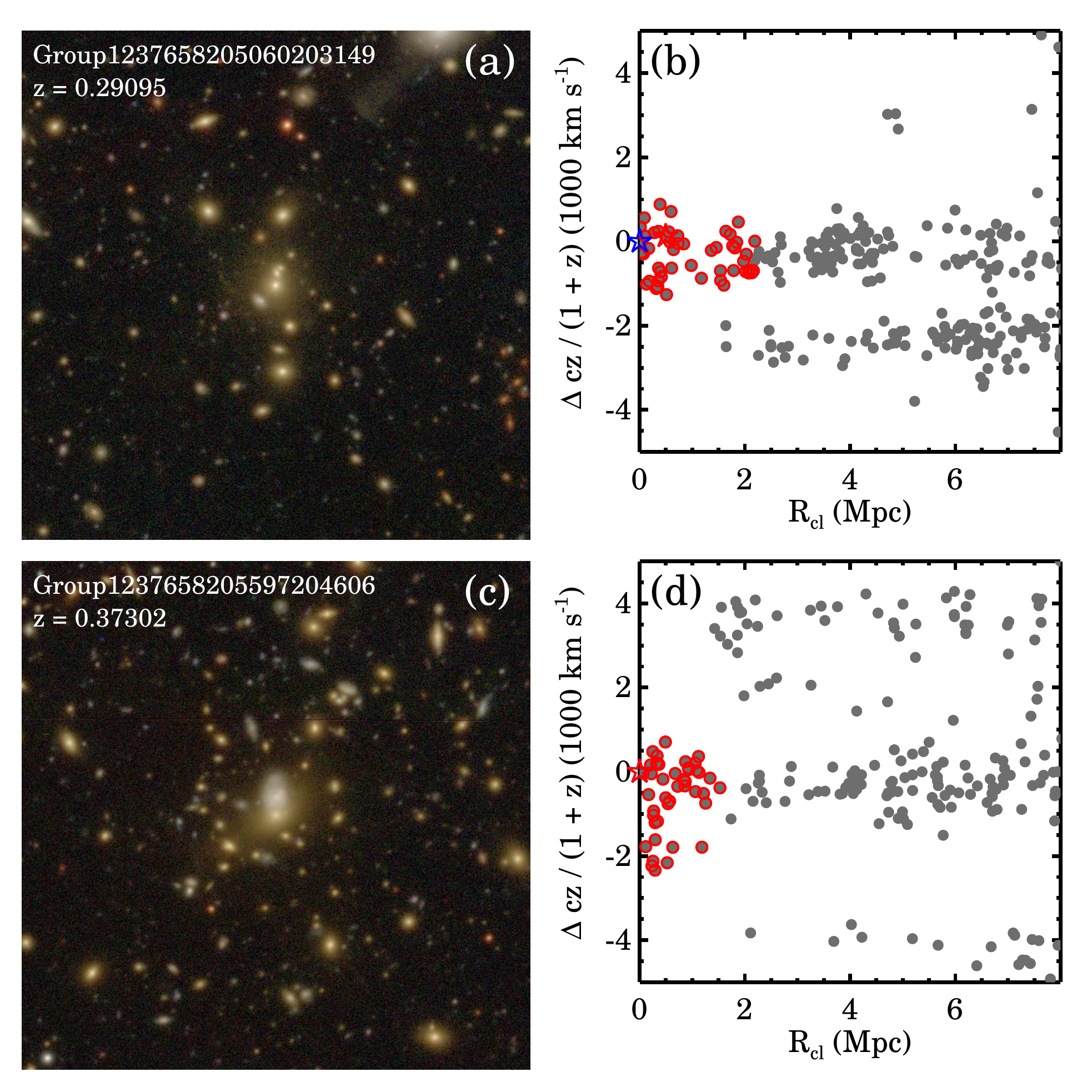}
\caption{(Left) Subaru/HSC images of two FoF clusters without a RM counterpart in the intermediate density regime. (Right) The R-v diagrams of the two clusters. Gray circles show galaxies with a spectroscopic redshift and the red circles mark the FoF members. }
\label{noRM_ID}
\end{figure}

Figure \ref{noRM_ID} shows Subaru/HSC images and R-v diagrams of two example FoF clusters with intermediate density and without a RM counterpart. The cluster members show a strong concentration in the HSC images. In the R-v diagrams, there is clear elongation of the cluster members along the line-of-sight, the signature of a massive cluster. All of the other clusters within the intermediate density show a similarly strong concentration in the HSC images and elongation in the R-v diagram. Differences in the catalog at the fiducial mass limit of the RM catalog probably reflect error in the velocity dispersion (FoF catalog) and/or the error in the richness (RM). 

% ========================================================
% Figure \ref{fof_regime}
% ========================================================
% ========================================================

\begin{figure}
\centering
\includegraphics[scale=0.40]{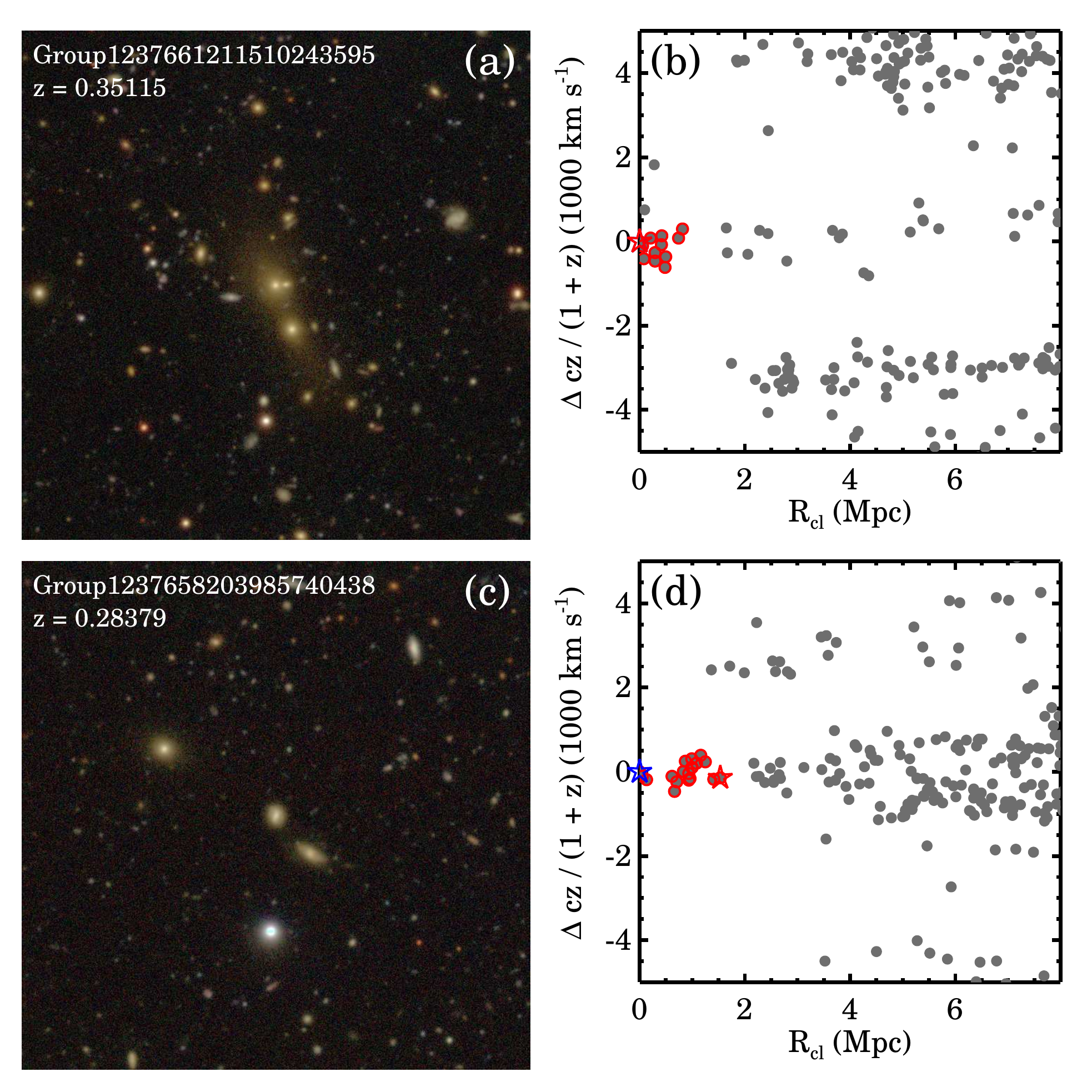}
\caption{Same as Figure \ref{noRM_ID}, but for two FoF clusters in the low-density regime without a redMaPPer counterpart. Panels (a) and (b) show a genuine cluster we identify based on the FoF algorithm. Panels (c) and (d) display an example of false positive where clustering around the central galaxy is weak. }
\label{noRM_LD}
\end{figure}

In the low-density regime, only 32 clusters have a RM counterpart. This result is not surprising because these systems may have masses much lower than the richness limit of the redMaPPer catalog. Even in this regime, a large fraction of the systems seems to be genuine clusters. For example, Figure \ref{noRM_LD} (a) and (b) display the HSC image and the R-v diagram of an FoF system with $\rho_{VL} = 0.11$. This FoF system has a dominant BCG at the center surrounded by many quiescent galaxies. The FoF members cluster around the BCG and there is the expected elongation in the radial direction. However, some systems with low number density are apparent false positives. Figure \ref{noRM_LD} (c) and (d) show FoF systems with $\rho_{VL} = 0.12$. Although this system consists of 13 spectroscopic members, but clustering around the central galaxy is weak. In the R-v diagram, the members extend to a larger projected distance, but only a few members are within the central region. There is no elongation.

% ========================================================
% Figure \ref{fof_regime}
% ========================================================
% ========================================================

\section{HectoMAP FoF Cluster Catalog}\label{sec:cat}

The HectoMAP FoF cluster catalog includes 346 clusters with 10 or more spectroscopic members. We list the properties of HectoMAP FoF clusters including R.A., Decl., central redshift, the number of FoF members, the cluster velocity dispersion, the galaxy number density, and the density flag in Table \ref{cat:cl}. The density flag indicates the density regime that includes the FoF cluster. In the high- and intermediate-density regime, there are no false positive FoF clusters. In the low-density regime, there are some likely false positives among FoF systems (see Section \ref{sec:explore}). We also list the FoF cluster members in Table \ref{cat:mem} including the FoF cluster ID, SDSS object ID, R.A., Decl, and redshift of the individual FoF members. 

% ========================================================
%Table \ref{cat:cl}
% ========================================================
\begin{deluxetable*}{lccccccccc}
\label{cat:cl}
\tablecaption{FoF Clusters in HectoMAP}
\tablecolumns{10}
\tabletypesize{\scriptsize}
\tablewidth{0pt}
% \tablehead{
% \colhead{ID} & \colhead{R.A.} & \colhead{Decl.} & \colhead{z} & \colhead{$N_{FoF, mem}^{1}$} & \colhead{$N_{spe, mem}^{2}$} & \colhead{$\sigma^{3}$} & \colhead{$\rho_{VL}$} & \colhead{$\rho_{VL}$ Flag$^{4}$} & \colhead{BCG Flag$^{5}$} }
\tablehead{
\multirow{2}{*}{ID}  & \colhead{R.A.}  & \colhead{Decl.}  & \multirow{2}{*}{z} &  
\multirow{2}{*}{$N_{FoF, mem}$\tablenotemark{$^{a}$}} &
\multirow{2}{*}{$N_{spe, mem}$\tablenotemark{$^{b}$}} &
\colhead{$\sigma^{c}$} & \colhead{$\rho_{VL}$} &
\multirow{2}{*}{$\rho_{VL}$ Flag$^{d}$} & 
\multirow{2}{*}{BCG Flag$^{e}$} \\
                     & \colhead{(deg)} & \colhead{(deg)}  &                    & 
                                                    &
                                                    &
\colhead{($\kms$)} & \colhead{(Mpc$^{-3}$)} & 
                                  &
                                  }
\startdata
HMRM001 & 200.633513 & 43.008366 &  0.281784 &  17 &  24 & $223 \pm  50$ & 0.15 & I & Y \\
HMRM002 & 203.101095 & 42.595151 &  0.304925 &  24 &  27 & $310 \pm  39$ & 0.12 & L & N \\
HMRM003 & 201.696117 & 43.188272 &  0.143471 &  10 &  26 & $386 \pm  74$ & 0.07 & L & Y \\
HMRM004 & 200.307439 & 43.506075 &  0.316042 &  12 &  14 & $378 \pm  60$ & 0.10 & L & Y \\
HMRM005 & 201.970415 & 43.264792 &  0.372521 &  10 &  10 & $288 \pm  71$ & 0.11 & L & Y \\
HMRM006 & 202.311345 & 43.232747 &  0.332000 &  17 &  20 & $514 \pm 110$ & 0.14 & L & Y \\
HMRM007 & 204.656461 & 42.817525 &  0.432667 &  11 &  11 & $400 \pm 154$ & 0.09 & L & Y \\
HMRM008 & 200.496818 & 43.173660 &  0.326603 &  11 &  16 & $166 \pm  31$ & 0.10 & L & Y \\
HMRM009 & 204.491698 & 42.824992 &  0.303717 &  25 &  32 & $422 \pm  72$ & 0.17 & I & Y \\
HMRM010 & 201.870106 & 43.083619 &  0.373789 &  11 &  11 & $514 \pm 146$ & 0.12 & L & Y \\
\enddata 
\tablenotetext{a}{Number of FoF members.}
\tablenotetext{b}{Number of spectroscopic members including galaxies fainter than $M_{r} = -19.72$.}
\tablenotetext{c}{Velocity dispersion. }
\tablenotetext{d}{The flag indicates the density regime: `H' is high-density regime, `I' is intermediate-density regime, and `L' is low-density regime. }
\tablenotetext{e}{The flag indicates that a central galaxy is the brightest cluster member. }
\end{deluxetable*}
% ========================================================

% ========================================================
%Table \ref{cat:mem}
% ========================================================
\begin{deluxetable*}{lccccc}
\label{cat:mem}
\tablecaption{HectoMAP FoF Cluster Members}
\tablecolumns{6}
\tabletypesize{\scriptsize}
\tablewidth{0pt}
\tablehead{
\colhead{Cluster ID} & \colhead{Object ID} & \colhead{R.A.} & \colhead{Decl.} & \colhead{z}} 
\startdata
HMRM001 & 1237661849863782556 & 200.633513 &  43.008366 & $0.28178 \pm 0.00007$ \\
HMRM001 & 1237661849863782610 & 200.607216 &  43.005065 & $0.28273 \pm 0.00011$ \\
HMRM001 & 1237661849863782898 & 200.683745 &  42.999933 & $0.28211 \pm 0.00009$ \\
HMRM001 & 1237661849863782555 & 200.638060 &  43.005394 & $0.28275 \pm 0.00009$ \\
HMRM001 & 1237661849863782459 & 200.582142 &  43.033193 & $0.28387 \pm 0.00010$ \\
HMRM001 & 1237661849863782755 & 200.605014 &  43.015250 & $0.28312 \pm 0.00017$ \\
HMRM001 & 1237661849863782558 & 200.648308 &  42.990242 & $0.28222 \pm 0.00009$ \\
HMRM001 & 1237661849863782826 & 200.654731 &  43.042562 & $0.28258 \pm 0.00022$ \\
HMRM001 & 1237661849863782753 & 200.605855 &  43.020831 & $0.28133 \pm 0.00012$ \\
HMRM001 & 1237661849863782894 & 200.680854 &  42.998204 & $0.28245 \pm 0.00021$ \\
\enddata 
\end{deluxetable*}
% ========================================================

\begin{figure*}[ht]
\centering
\includegraphics[scale=0.44]{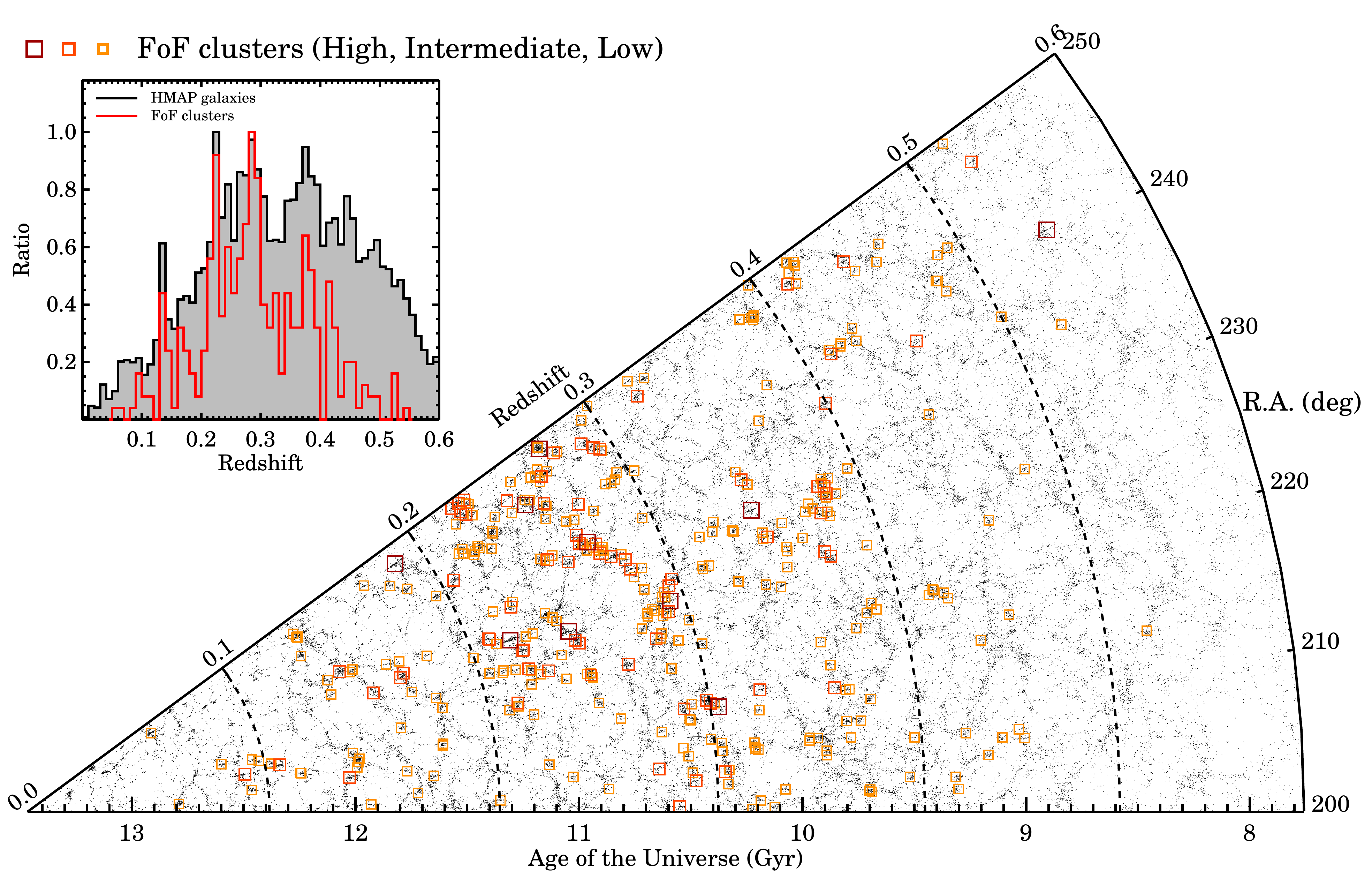}
\caption{HectoMAP cone diagram projected in R.A.. Black points show all HectoMAP galaxies with spectroscopic redshifts. Squares mark the HectoMAP FoF clusters; darker and larger symbols indicate higher density. The inset image shows the normalized redshift distribution of the FoF clusters (red open histogram) and the full HectoMAP survey (black filled histogram).}
\label{cone}
\end{figure*}

The cone diagram in Figure \ref{cone} shows the distribution of spectroscopic objects and FoF clusters. Squares mark the location of the FoF clusters; the darker and larger symbols indicate higher density. The FoF systems follow the large scale structure defined by all of galaxies in HectoMAP. The inset image shows the FoF cluster redshift distribution (red histogram). For comparison, we also plot the redshift distribution of the entire HectoMAP survey. At $z > 0.35$, the sampling in HectoMAP only enables identification of dense, relatively massive systems. 

% ========================================================
% Figure \ref{fof_summary}
% ========================================================
% ========================================================

We next explore the BCG properties of the HectoMAP FoF clusters. The BCG is a distinctive galaxy often located at the bottom of the cluster potential well. Because we identify the BCGs based on the center of light method, which takes the luminosity density around the central galaxy into account, the BCGs of the FoF clusters are generally close to the center of the cluster and identification is obvious. However, in this process the BCG identification can be confused because of uncertainties in galaxy photometry in the crowded central region or because of contamination by other bright galaxies in the outskirts of the cluster (e.g., \citealp{Sohn19}). Among 346 FoF clusters, there are 86 systems ($\sim 25\%$) where the brightest cluster member is not identical to the object identified by the center of light method. 

\begin{figure}
\centering
\includegraphics[scale=0.35]{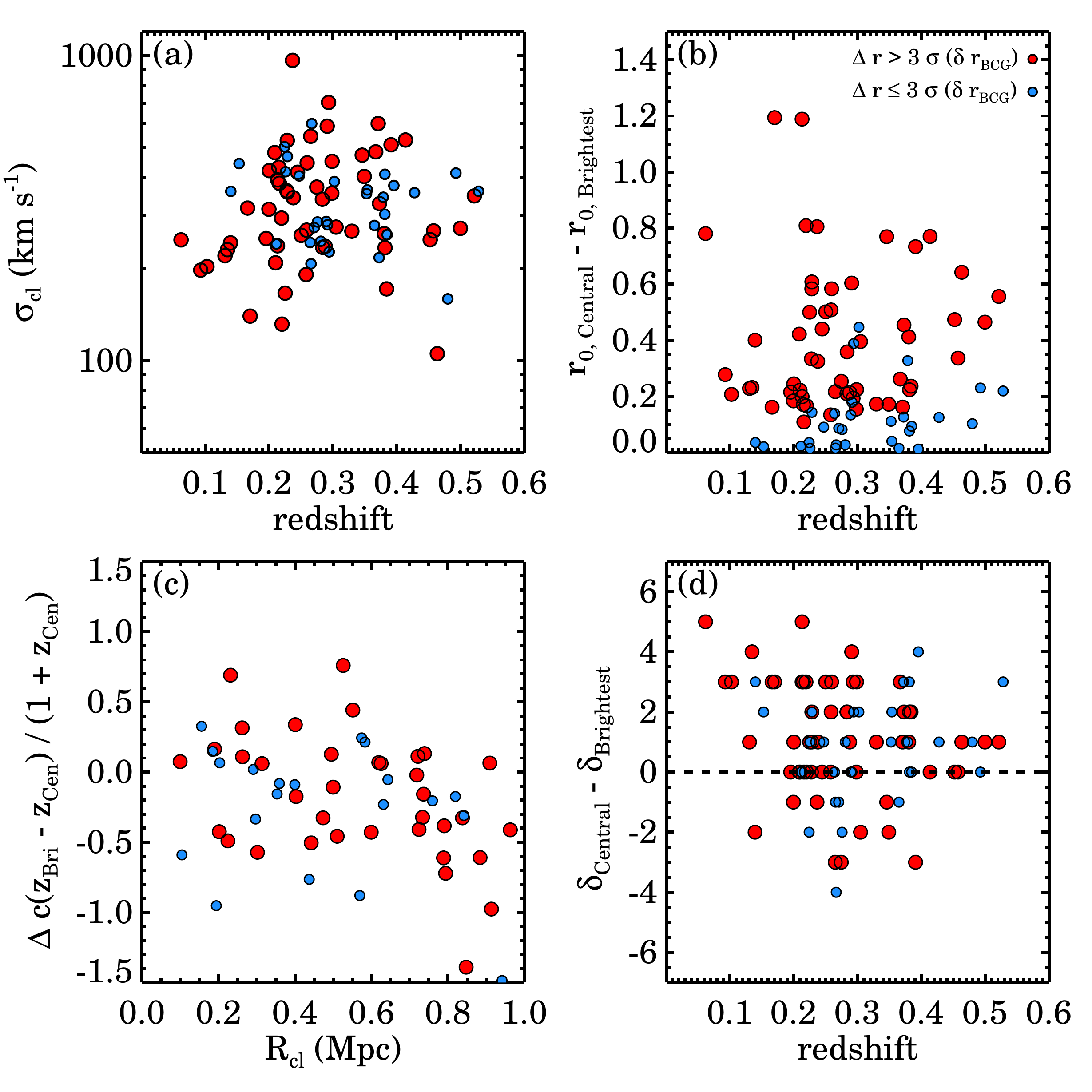}
\caption{(a) Velocity dispersion vs. redshift for 86 FoF clusters where the brightest cluster member is not identical to the central galaxy identified by the center of light method. Blue (and red) circles indicate  FoF systems where the magnitude difference is less (and more) than $3\sigma$ uncertainty in the BCG photometry (see panel (b)). (b) The magnitude difference between the central galaxy and the brightest cluster member as a function of cluster redshift. (c) The stacked R-v diagram of the 86 brightest members in FoF systems where the brightest cluster member is not the central galaxy. The projected distance and the relative radial velocity of the brightest cluster members are computed with respect to the central galaxy. (d) The difference between the local density around the central galaxy and the brightest cluster member as a function of cluster redshift. }
\label{bcg_nocentral}
\end{figure}

Figure \ref{bcg_nocentral} (a) shows the velocity dispersion of these 86 FoF systems as a function of redshift. The Kolmogorov-Smirnov test suggests that the distributions of the redshift and velocity dispersion of these 86 systems are not significantly different from the full sample with a significance level of 0.05 and 0.47, respectively. 

Figure \ref{bcg_nocentral} (b) shows the magnitude difference between the central galaxy and the brightest cluster member ($\Delta r = r_{0, Central} - r_{0, Brightest}$) in the 86 FoF clusters where the choice of the BCG is not obvious. In 30 systems ($\sim35\%$), the magnitude difference is less than the $3\sigma$ uncertainty in the BCG magnitude. In other words, the BCG identification can be confused because of the large uncertainty in the photometry. In these cases, the central and the brightest galaxies often have nearby companions that affect the galaxy photometry. In the other 56 systems, the brightest members are brighter than the central galaxies by $0.1 - 1.3$ mag. 

Figure \ref{bcg_nocentral} (c) shows the relative velocity difference and the projected distance between the brightest member and the central galaxies in 86 problematic FoF systems. The brightest members are located at $0.1 < R_{cl} ({\rm Mpc}) < 1.0$ and $|\Delta cz / (1 + z_{cl})| < 1000~\kms$. The stacked R-v diagram shows that the brightest members in these cases are actually in the cluster outskirts. 

Figure \ref{bcg_nocentral} (d) displays the difference between the local number density around the central galaxy and around the brightest member as a function of cluster redshift. Here, the local density is the galaxy number count within a cylindrical volume with $R_{proj} < 200$ kpc and $|\Delta cz /(1 + z_{cl})| < 1300~\kms$. A positive local density difference indicates that the local density around the central galaxy exceeds that around the brightest cluster member. In a majority of the systems ($\sim 83\%$), the local density difference is positive, suggesting that the central galaxy is a better BCG choice because it sits nearer to the potential minimum. There are only 8 systems where the local density around the brightest cluster member exceeds the density around the central galaxy and where the magnitude difference is significant. 

In conclusion, the central galaxies we identify are indeed brightest cluster galaxies (BCGs) in a majority ($\sim 75\%)$ of the HectoMAP FoF clusters. In 86 systems, the BCG identification is confused by the brighter galaxies located in the cluster outskirts. We mark these 86 systems in Table \ref{cat:cl}. For further discussion, we include the central galaxies in these 86 clusters. Excluding these clusters does not impact the results of the analysis. 

% ========================================================
% Figure \ref{lens_candidate}
% ========================================================
% ========================================================

\begin{figure*}
\centering
\includegraphics[scale=0.36]{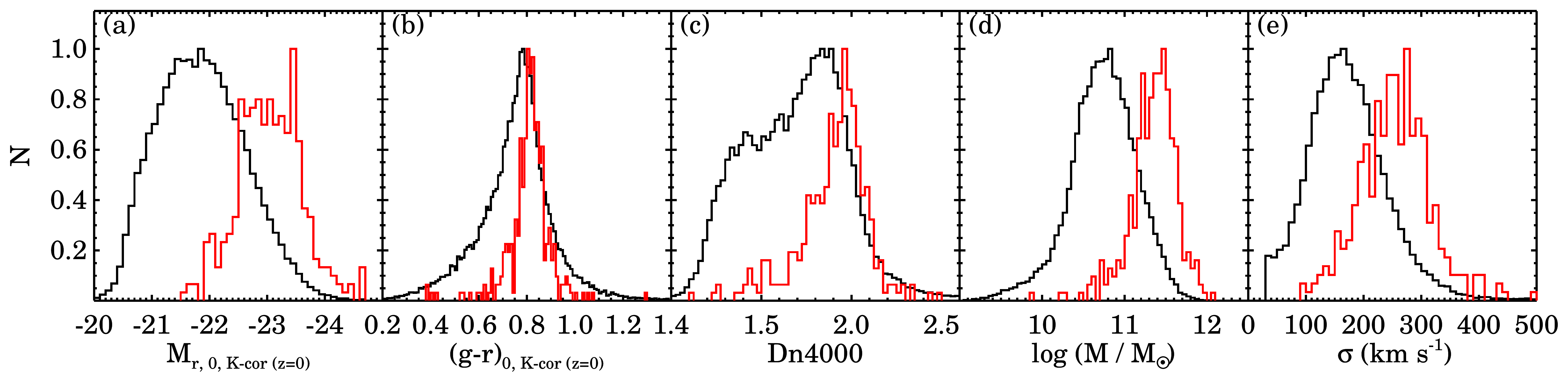}
\caption{Distributions of the physical properties of the BCGs (red open histograms) including (a) foreground extinction and K-corrected $r-$band absolute magnitudes, (b) $(g-r)$ color, (c) $\dn$, (d) stellar mass, and (e) stellar velocity dispersion. For comparison, black filled histograms display the distributions of all FoF cluster members.}
\label{bcg_prop}
\end{figure*}

We next examine the internal physical properties of the BCGs in all 346 HectoMAP FoF clusters. Figure \ref{bcg_prop} displays the distributions of the physical properties of the BCGs, including (a) foreground extinction and K-corrected $r-$band absolute magnitudes, (b) $(g-r)$ color, (c) $\dn$, (d) stellar mass, and (e) stellar velocity dispersion. For comparison, we also plot the same distributions for all of the FoF cluster members (black histograms). Figure \ref{bcg_prop} demonstrates that the BCGs are a distinctive population. The $\dn$ distribution shows that the most of BCGs ($\sim 96\%$) are quiescent ($\dn > 1.5$). The HectoMAP BCGs are very massive; $\sim 90\%$ of the BCGs have $\log (M_{*} / M_{\odot}) > 11$. The stellar velocity dispersions of the BCGs are also generally large compared to those of other cluster galaxies, although the range is quite broad ($100 < \sigma~(\kms) < 450$). 

% ========================================================
% Figure \ref{lens_candidate}
% ========================================================
% ========================================================

We list the physical properties of the BCGs in Table \ref{cat:bcg}. We include the SDSS object ID, foreground extinction and K-corrected $r-$band absolute magnitude, $(g-r)$ color, $\dn$, and the central stellar velocity dispersion of the BCGs ($\sigma_{*, BCG}$). Here, inclusion of $\sigma_{*, BCG}$ is a unique feature of the HectoMAP FoF cluster sample. We explore the relation between the physical properties of the BCGs and the FoF clusters in Section \ref{sec:connection}. 

%=================================
%Table \ref{BCG_cat}
%=================================
\begin{deluxetable*}{llcccc}
\label{cat:bcg}
\tablecaption{BCGs of in the HectoMAP FoF Clusters}
\tablecolumns{6}
\tabletypesize{\scriptsize}
\tablewidth{0pt}
\tablehead{
\colhead{Cluster ID} & \colhead{BCG Object ID} & \colhead{$M_{r}$} & \colhead{$(g-r)$} & \colhead{$\dn$} & \colhead{$\sigma$}}
\startdata
HMRM001 & 1237661849863782556 & $-22.30 \pm 0.02$ & 1.71 & $2.13 \pm 0.05$ & $273 \pm 13$ \\
HMRM002 & 1237661849864568984 & $-21.97 \pm 0.06$ & 1.64 & $2.12 \pm 0.11$ & $182 \pm 24$ \\
HMRM003 & 1237661850400981101 & $-21.98 \pm 0.02$ & 1.63 & $1.72 \pm 0.03$ & $199 \pm 12$ \\
HMRM004 & 1237661850400522409 & $-22.40 \pm 0.03$ & 1.72 & $1.87 \pm 0.08$ & $273 \pm 23$ \\
HMRM005 & 1237661850401046742 & $-22.31 \pm 0.05$ & 1.65 & $1.91 \pm 0.05$ & $233 \pm 20$ \\
HMRM006 & 1237661850401177769 & $-22.10 \pm 0.07$ & 1.78 & $2.06 \pm 0.06$ & $332 \pm 23$ \\
HMRM007 & 1237661850401964328 & $-22.42 \pm 0.08$ & 1.76 & $1.96 \pm 0.13$ & $266 \pm 32$ \\
HMRM008 & 1237661871871623348 & $-22.18 \pm 0.04$ & 1.71 & $1.75 \pm 0.03$ & $184 \pm 13$ \\
HMRM009 & 1237661850401898642 & $-22.54 \pm 0.03$ & 1.69 & $2.23 \pm 0.06$ & $277 \pm 17$ \\
HMRM010 & 1237661871872082077 & $-23.26 \pm 0.05$ & 1.85 & $2.13 \pm 0.13$ & $339 \pm 24$ \\
\enddata 
\end{deluxetable*}
%=================================

\section{CONNECTION BETWEEN BRIGHTEST CLUSTER GALAXIES AND CLUSTERS}\label{sec:connection}

Dense spectroscopy of galaxy clusters enables interesting dynamical analyses that connect clusters with their BCGs. \citet{Sohn20} demonstrate the application of dense spectroscopy to explore the connection between clusters and their BCGs. 

The large HectoMAP cluster catalog not only doubles the sample size of \citet{Sohn20} for exploring this relation, but it also provides a sample that covers wider redshift ($0.1 < z < 0.6$) and mass ranges ($100 < \sigma_{cl} (\kms) < 1000$). This redshift range is important because clusters double their mass from a redshift $0.5 - 0.6$ to the present (e.g., \citealp{Fakhouri10, Haines18, Pizzardo21}). BCGs develop in tandem with their host clusters (e.g., \citealp{DeLucia07}). Exploring lower mass systems also provides a more extensive picture of the relationship between clusters and their central galaxies.  

In \citet{Sohn20}, we investigate the relationship between the BCG stellar velocity dispersion (hereafter $\sigma_{*, BCG}$) and the cluster velocity dispersion (hereafter $\sigma_{cl}$). \citet{Sohn20} use the HeCS-omnibus sample that compiles spectroscopic data for 223 massive clusters. HeCS-omnibus includes clusters at $0.02 < z < 0.29$ with a median redshift of $0.10$. The masses of the HeCS-omnibus clusters range from $2.5 \times 10^{14} M_{\odot}$ to $1.8 \times 10^{15} M_{\odot}$ with a median mass of $3.0 \times 10^{14} M_{\odot}$, corresponding to $210 < \sigma_{cl}~(\kms) < 1350$ with a median $\sigma_{cl}$ of $\sim 700~\kms$. The HeCS-omnibus clusters are the most massive clusters selected from a large volume that covers almost half of the sky (i.e., the northern hemisphere). The HectoMAP cluster sample potentially probes the evolution of the relation between the median redshift of HeCS-omnibus and HectoMAP.

% ========================================================
% Figure \ref{bcg_ratio}
% ========================================================
\begin{figure}
\centering
\includegraphics[scale=0.35]{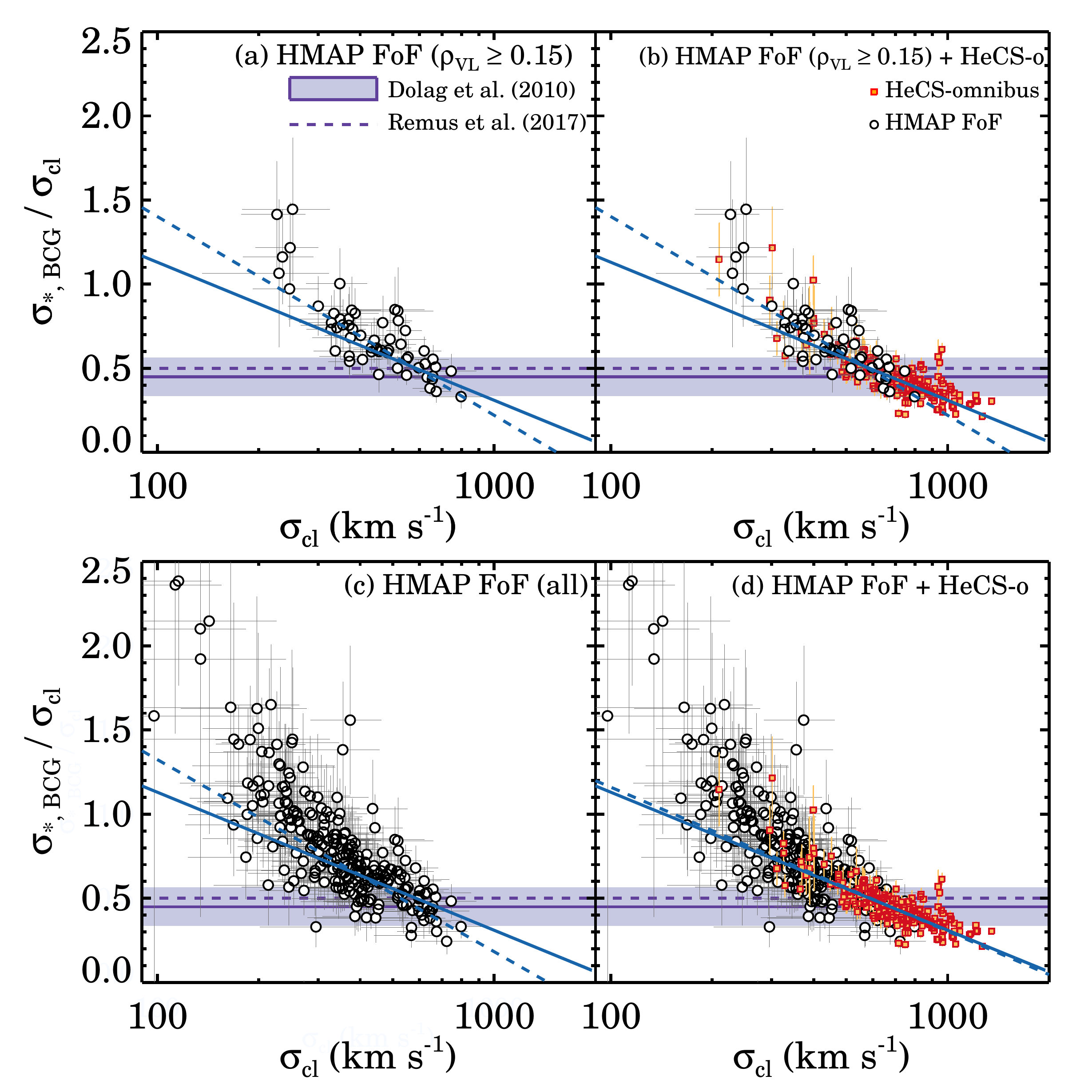}
\caption{Ratio between the stellar velocity dispersion of BCGs and the cluster velocity dispersion ($\sigma_{*, BCG}/\sigma_{cl}$) vs. cluster velocity dispersion ($\sigma_{cl}$) for (a) 80 HectoMAP FoF clusters with $\rho_{cl} > 0.15$, (b) 80 HectoMAP FoF clusters and 223 HeCS-omnibus clusters, (c) the full HectoMAP FoF clusters, and (3) the full HectoMAP and HecS-omnibus clusters. Black circles and red squares are HectoMAP FoF and HeCS-omnibus clusters. The blue solid line shows the best-fit relation derived from the HeCS-omnibus clusters \citep{Sohn20}. The blue dashed line shows the best-fit relation we derive for each subsample in the panel. The purple horizontal solid line and the shaded region show the theoretical prediction and its $1\sigma$ boundary from \citet{Dolag10}. The horizontal dashed line indicates a similar prediction from \citet{Remus17}. }
\label{bcg_ratio}
\end{figure}
% ========================================================

Figure \ref{bcg_ratio} (a) shows the ratio between $\sigma_{*, BCG}$ and $\sigma_{cl}$ for 80 HectoMAP FoF clusters as a function of $\sigma_{cl}$. Here, we plot only FoF clusters within the high- and intermediate density regime; all of these systems are genuine massive clusters. We note that all 80 clusters have quiescent BCGs (i.e., $\dn > 1.5$). Thus, the stellar velocity dispersion of the BCGs is a good mass proxy. These 80 HectoMAP clusters show a remarkably tight relation; the ratio declines as a function of $\sigma_{cl}$. We use a Markov Chain Monte Carlo (MCMC) approach to derive the best-fit relation for these clusters (the blue dashed line): 
\begin{equation}
\sigma_{*, BCG} / \sigma_{cl} = (-1.12 \pm 0.14) \log \sigma_{cl} + (3.60 \pm 0.37).
\end{equation}

We compare the relation from HectoMAP clusters with the best-fit relation for HeCS-omnibus clusters (the blue solid line in Figure \ref{bcg_ratio}, \citealp{Sohn20}): 
\begin{equation}
\sigma_{*, BCG} / \sigma_{cl} = (-0.82 \pm 0.17) \log \sigma_{cl} + (2.77 \pm 3.93).
\end{equation}
In Figure \ref{bcg_ratio} (b), red squares show 223 HeCS-omnibus clusters. Because HeCS-omnibus includes $\sim180$ spectroscopic members in each cluster, individual cluster velocity dispersions have much small uncertainties ($\lesssim 50~\kms$). The relation derived for the HeCS-omnibus sample is slightly shallower than the relation of the HectoMAP dense clusters, but the difference is not significant ($< 2.1\sigma$). 

Figure \ref{bcg_ratio} (c) displays the same relation for the full HectoMAP FoF sample. In this plot, we use 331 FoF clusters with quiescent BCGs. We exclude 15 clusters that host non-quiescent BCGs, because the velocity dispersion of these BCGs could be dominated by ordered rotation of the disk. Remarkably, the relation between $\sigma_{*, BCG}/\sigma_{cl}$ and $\sigma_{cl}$ is tight for $\sigma_{cl} \gtrsim 250~\kms$: 
\begin{equation}
\sigma_{*, BCG} / \sigma_{cl} = (-1.06 \pm 0.07) \log \sigma_{cl} + (3.39 \pm 0.19).
\end{equation}

We compare the full HectoMAP FoF and HeCS-omnibus samples in Figure \ref{bcg_ratio} (d). In general, the relationship derived from the two independent cluster samples at different redshifts and covering different mass ranges is striking. We derive the best-fit relation based on the combined HectoMAP and HeCS-omnibus sample: 
\begin{equation}
\sigma_{*, BCG} / \sigma_{cl} = (-0.82 \pm 0.03) \log \sigma_{cl} + (2.77 \pm 0.09).
\end{equation}
This relation is essentially identical to the relation based only on the HeCS-omnibus sample because of the small uncertainties in the HeCS-omnibus velocity dispersions. 

\section{DISCUSSION}\label{sec:discussion}

The tight observed relation between $\sigma_{*, BCG}$ and $\sigma_{cl}$ suggests an interesting evolutionary scenario for BCGs and their host clusters. The relation indicates that the mass fraction associated with the BCG changes as a function of cluster mass. In massive clusters, the mass associated with the BCGs decreases steadily relative to the cluster mass. In low mass systems, the BCG mass is comparable with the cluster mass. 

Here, we discuss the implication of the $\sigma_{*,BCG}/\sigma_{cl} - \sigma_{cl}$ relation. We compare the observed relation with the prediction from numerical simulations in Section \ref{sec:sim}. We then explore the plausible redshift evolution of the relation in Section \ref{sec:evolution}. We discuss a possible evolutionary scenario of the BCGs based on this relation in Section \ref{sec:scenario}.

\subsection{Comparison with Numerical Simulations}\label{sec:sim}

We compare the observed relation between $\sigma_{*, BCG}/\sigma_{cl}$ and $\sigma_{cl}$ with  predictions from numerical simulations. We use the results from \citet{Dolag10} because they include stellar velocity dispersion measurements that can be compared with the observations. \citet{Dolag10} explore the relation between the BCG and cluster velocity dispersions based on numerical simulations that include 44 clusters with $M_{200} > 5 \times 10^{13} M_{\odot}$ (or $\sigma_{cl} \gtrsim 300~\kms$). They identified star particles that are not bound to any subhalos within the cluster potential. These star particles show a two component velocity distribution; one component belongs to the BCG (cD galaxy) central potential and another one is associated with the diffuse stellar halo (DSC). They compute the velocity dispersions of these two components as $\sigma_{BCG}$ and $\sigma_{DSC}$ (i.e., $\sim \sigma_{cl}$). 

Interestingly, the ratio between $\sigma_{BCG}$ and $\sigma_{cl}$ measured from the simulations is independent of $\sigma_{cl}$. \citet{Dolag10} show that both $\sigma_{BCG}$ and $\sigma_{cl}$ are well correlated with the cluster halo mass ($M_{halo} \sim \sigma^{3}$), and thus the ratio between $\sigma_{BCG}$ and $\sigma_{cl}$ remains constant: $\sigma_{*, BCG} \simeq (0.45 \pm 0.11) \sigma_{cl}$. In Figure \ref{bcg_ratio}, the horizontal solid line and the shaded region mark the relation derived from \citet{Dolag10} and the $1\sigma$ range. \citet{Remus17} derived a similar relation based on simulations with higher resolution and a larger box size: $\sigma_{*, BCG} = 0.5 \sigma_{cl}$ (the dashed line). 

The discrepancy between the observed and theoretical relations for $\sigma_{*, BCG}$ and $\sigma_{cl}$ offers an intriguing test of BCG and cluster formation models. Many previous studies explore the evolution of BCGs based primarily on BCG stellar mass, which is sensitive to complex baryonic physics (e.g., feedback models) in numerical simulations. Observed stellar mass estimates  are affected by photometric uncertainties in the crowded cluster core and by systematic biases introduced by the choice of stellar population model. The central stellar velocity dispersion  is insensitive to systematic observational biases and is relatively straightforward to measure. In future simulations, the velocity dispersion of the BCG could be measured based on particles within a cylindrical region that penetrates the central region of the BCG for more direct comparison with the spectroscopic observations (e.g., \citealp{Zahid18}). 

\subsection{Tracing the Coevolution of BCGs and Their Host Clusters}\label{sec:evolution}

We next explore BCGs and their host cluster at different redshifts. We select 78 HeCS-omnibus clusters with $z < 0.1$ and 97 HectoMAP FoF clusters with $0.3 < z < 0.4$. The age difference between these two redshift epochs is $\sim 3$ Gyrs. The HeCS-omnibus subsample includes very massive systems with $210 < \sigma_{cl} (\kms) < 963$ with a median $\sigma_{cl} = 622~\kms$. In contrast, the HectoMAP subsample includes generally lower velocity dispersion systems ($126 < \sigma_{cl} (\kms) < 797$ with a median $\sigma_{cl} = 353~\kms$). The HeCS-omnibus clusters at low redshift tend to be more evolved systems with large mass. The selection of the HectoMAP and HeCS-omnibus samples differ substantially. HectoMAP is a comprehensive FoF sample in its redshift range; HeCS-omnibus collects available  data from the literature. In spite of the differences in catalog construction, the two samples provide a baseline for comparing sets of clusters are different epochs.

\begin{figure}
\centering
\includegraphics[scale=0.5]{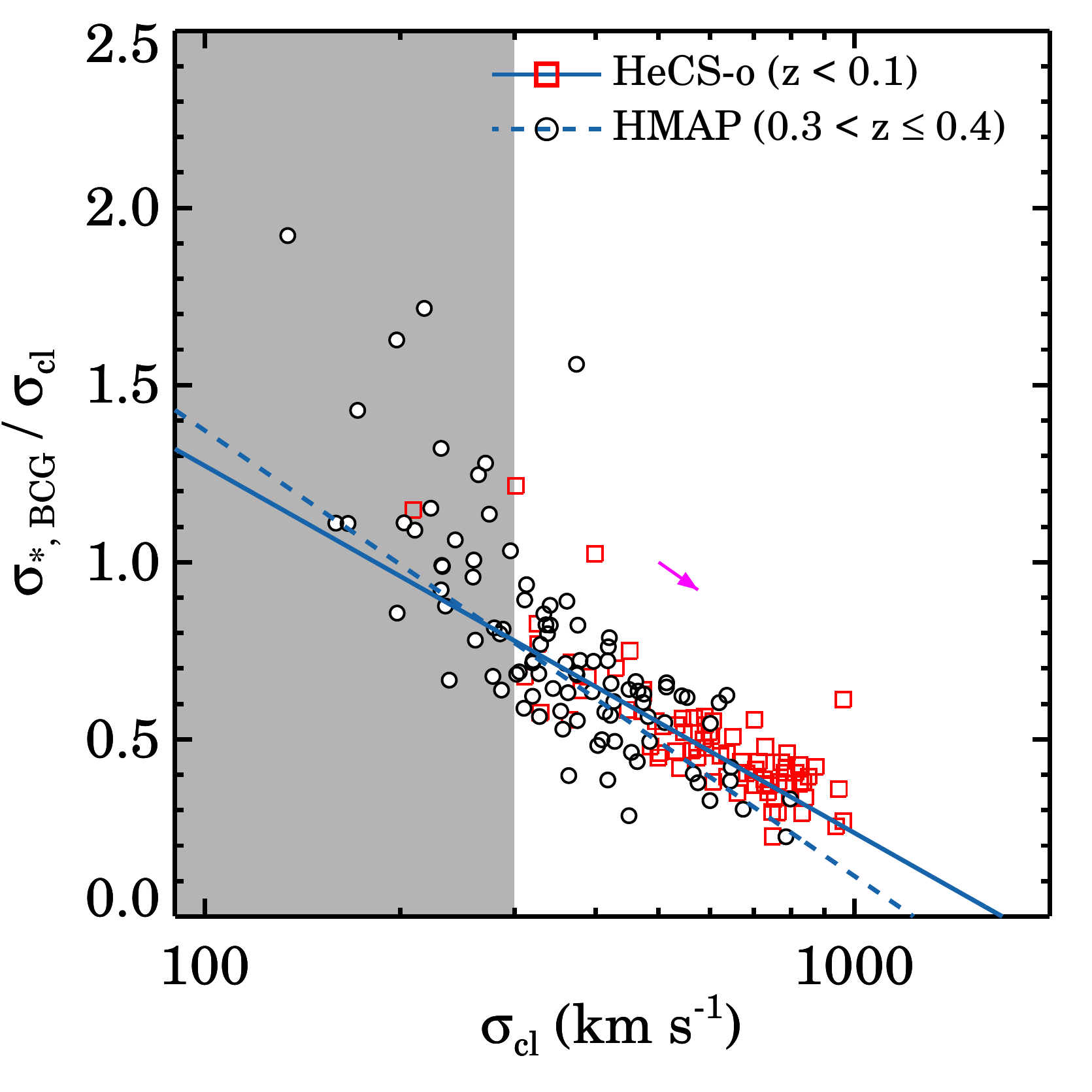}
\caption{Same as Figure \ref{bcg_ratio}, but for subsamples from the HeCS-omnibus (red squares) and HectoMAP FoF cluster catalogs (black circles). The solid and dashed lines show the best-fit for HeCS and HectoMAP, respectively. The shaded region indicates where the clusters deviate from the relation. The magenta arrow marks the expected evolutionary direction based on simulated clusters combined with minor mergers (e.g., \citealp{Haines18}.) }
\label{comp_ratio}
\end{figure}

Figure \ref{comp_ratio} shows the $\sigma_{*, BCG}/\sigma_{cl} - \sigma_{cl}$ relation for the HectoMAP (black circles) and HeCS-omnibus (red squares) subsamples. The slopes of the best-fit relations are consistent: $(-1.04 \pm 0.10)$ for the HeCS-omnibus and $(-1.26 \pm 0.25)$ for the HectoMAP. These slopes are based on subsamples with $\sigma_{cl} < 800~\kms$, the maximum $\sigma_{cl}$ of the HectoMAP subsample. The slope for the full HeCS-omnibus subsample is slightly shallower ($-0.84 \pm 0.09)$, but within $2\sigma$ of the HectoMAP sample. 

The remarkable consistency in slope indicates that the ratio between the BCG and the cluster mass evolves along the relation in Figure \ref{comp_ratio} as the universe ages over the last $\sim 3$ Gyrs. The HectoMAP systems at $z \sim 0.35$ presumably evolve into more massive clusters (e.g., \citealp{Zhao09, Fakhouri10, Haines18}). More specifically, if a cluster halo reaches a velocity dispersion of $\sigma_{cl} \sim 300~\kms$, the BCG mass growth is slower than the growth of the clsuter halo. 

The direction of the arrow in Figure \ref{comp_ratio} assumes the cluster growth rate from \citet{Haines18} and a negligible change in the BCG velocity dispersion as might be expected if BCG growth is dominated by minor mergers at these epochs (e.g., \citealp{Edwards20}). The arrow essentially parallels the observations. The length of the arrow indicates the expected magnitude of the evolution; the predicted length spanning these two epochs is a change in cluster velocity dispersion of about $\sim 80~\kms$. Based on shells at large radius for the HectoMAP cluster centers (Pizzardo et al. 2021, in preparation) show that cluster growth in the HectoMAP sample is consistent with the prediction of these simulations.

% ========================================================
% Figure \ref{bcg_ratio}
% ========================================================
% ========================================================

\subsection{Growth Mechanisms for the BCG }\label{sec:scenario}

The tight relation in Figure \ref{bcg_ratio} and Figure \ref{comp_ratio} indicates that the mass fraction associated with BCG decreases continuously as a function of cluster mass (at $\sigma_{cl} > 300~\kms$). The slope of the relation suggests that the BCG mass growth is slow over the redshift range we explore. BCG growth in massive clusters seems to be  slower than in less massive systems. 

The apparently slow growth of BCGs supports the idea that at late times minor mergers and/or accretion of stripped material (e.g., \citealp{Contini18, RagoneFigueroa18}) are the dominant mechanism for BCG growth. High mass clusters have a large velocity dispersion that precludes major mergers.  

We note that the relation for HectoMAP clusters in the shaded region in Figure \ref{comp_ratio} steepens at $\sigma_{cl} < 300~\kms$. The apparent steepening of the relation is even more evident for the full HectoMAP sample (Figure \ref{bcg_ratio}). The impact of major mergers can affect BCG growth in this velocity dispersion range where the cluster and BCG dispersions are similar. Major mergers lead to significant increases in the central velocity dispersion of the resultant object (e.g., \citealp{Hilz12}). Interestingly, some HectoMAP FoF systems with $\sigma_{cl} < 300~\kms$ host BCGs that show sign of recent mergers (e.g., shell structures) in the HSC images. The presence of mergers in the lower dispersion systems may be evidence of the role of pre-processing (e.g., \citealp{Balogh02, Fujita04}) in the development of BCGs. A systematic study of these BCGs will provide more insights into BCG evolution (Sohn et al. in preparation). A general picture of BCG growth is emerging where major mergers, minor mergers and accretion of stripped material all play a role (e.g., \citealp{Diaferio01, Lin04, DeLucia07, Laporte13, RagoneFigueroa18, Spavone21}), but the timing for each process is probably restricted by  local cluster dynamics.

\section{Conclusion}\label{sec:conclusion}

HectoMAP is a dense spectroscopic survey covering 54.64 deg$^{2}$ of the sky. A central goal of the HectoMAP redshift survey is identification of galaxy clusters based on spectroscopy and to explore the coevolution of the clusters and their members. In \citet{Sohn18a}, we use the HectoMAP survey to test the photometrically identified redMaPPer clusters. In \citet{Sohn18b}, we identify 15 X-ray clusters based on ROSAT all-sky X-ray data that are associated with the spectroscopic overdensities. Ultimately we plan to use the Subaru HSC imaging to measure weak lensing masses for the systems identified spectroscopically. eROSITA should soon provide X-ray masses throughout the mass and redshift range \citep{Merloni12, Predehl21}.

To build the catalog we apply a Friends-of-Friends (FoF) algorithm in redshift space. We use galaxies brighter than $M_{r} = -19.72$ in a volume-limited sample to $z = 0.35$ to determine linking lengths. We then extend these fiducial lengths them throughout the survey range. At redshifts $ z > 0.35$ the FoF catalog is dominated by relatively denser, more massive systems.  

The properties of FoF systems depend on the choice of linking lengths. We determine the linking lengths empirically based on comparison with redMaPPer clusters in the HectoMAP region. We test a set of projected and radial linking lengths, and find the optimal set of linking lengths (900 kpc and $500~\kms$) that recovers redMaPPer clusters (56/57 in the test sample). These linking lengths identify systems with a density larger than $~110$ times the typical density of the universe at a cluster redshift. 

The final HectoMAP FoF cluster catalog includes 346 systems with 10 or more spectroscopic members. We provide the FoF catalog including the membership, the BCG identification, and the BCG central stellar velocity dispersion. We divide the sample into three categories based on the galaxy number density around the cluster center. We investigate Subaru/HSC images and R-v diagrams of these systems. Systems in high- and intermediate-density regimes are all genuine clusters with strong concentration in the image and the elongation in the R-v diagram. In the high density regime all of the FoF clusters have RM counterparts; in the intermediate density regime, the FoF find 45\% more clusters than redMaPPer. In the low-density regime the FoF naturally includes some probable false positives ($\sim 30\%$) with no elongation in the R-v diagram.  

Based on the 346 FoF clusters, we explore the connection between the BCGs and their host clusters. Following \citet{Sohn20}, we investigate the relation between cluster velocity dispersion ($\sigma_{cl}$) and the stellar velocity dispersion of the BCGs ($\sigma_{*, BCG}$). The ratio between $\sigma_{*, BCG}$ and $\sigma_{cl}$ decreases as a function of $\sigma_{cl}$. This trend is consistent with the one for the HeCS-omnibus cluster sample \citep{Sohn20}. The slope of the relation is remarkably tight for $\sigma_{cl} > 300~\kms$ in both the HectoMAP (especially the high- and intermediate-density samples) and the HeCS-omnibus samples. 

In contrast with the data, numerical simulations predict a constant $\sigma_{*, BCG}/\sigma_{cl}$ ratio over a large $\sigma_{cl}$ range \citep{Dolag10, Remus17}. This discrepancy between the observed relation and the theoretical prediction offers an interesting test of coordinated BCG and cluster evolution. 

As a probe of the synergy between BCG and cluster evolution, we compare the $\sigma_{*, BCG}/\sigma_{cl} - \sigma_{cl}$ relation at two different redshifts based on HeCS-omnibus and HectoMAP. The relations from the two subsamples have the same slope, suggesting BCGs evolve along the relation as cluster accrete surrounding material. BCG evolution must be slow in massive clusters over the redshift range explored by HectoMAP. The data suggest that at late times BCGs in massive clusters ($\sigma_{cl} > 300~\kms$) grows mainly by minor mergers that produce a negligible increase in the BCG velocity dispersion. For systems with low velocity dispersion $\sigma_{*, BCG}/\sigma_{cl} - \sigma_{cl}$, an apparent steepening of the relation may result from the major mergers. 

The observational indications of the changing role of various BCG growth processes with velocity dispersion and possibly cosmic time can be tested with current high resolution simulations (e.g., \citealp{Springel18}). Additional observational constraints will obviously come from larger surveys and from multiple observational approaches to the HectoMAP FoF catalog including strong lensing, weak lensing, and X-ray observations. 

\acknowledgements
We thank Perry Berlind, Michael Calkins, and Nelson Caldwell for operating Hectospec. We thank Susan Tokarz, Jaehyon Rhee and Sean Moran for their significant ontributions to the data reduction. We also thank Scott Kenyon, Ivana Damjanov, Adi Zitrin, and Mark Vogelsberger for discussions that clarified the paper. J.S. is supported by the CfA Fellowship. M.J.G. acknowledges the Smithsonian Institution for support. H.S.H. is supported by the New Faculty Startup Fund from Seoul National University. A.D. acknowledges partial support from the INFN grant InDark and the Italian Ministry of Education, University and Research (MIUR) under the {\it Departments of Excellence} grant L.232/2016. This research has made use of NASA’s Astrophysics Data System Bibliographic Services.

Funding for the SDSS-IV has been provided by the Alfred P. Sloan Foundation, the U.S. Department of Energy Office of Science, and the Participating Institutions. SDSS-IV acknowl- edges support and resources from the Center for High Performance Computing at the University of Utah. The SDSS website is www.sdss.org. SDSS-IV is managed by the Astrophysical Research Consortium for the Participating Institutions of the SDSS Collaboration including the Brazilian Participation Group, the Carnegie Institution for Science, Carnegie Mellon University, Center for Astrophysics | Harvard and Smithsonian, the Chilean Participation Group, the French Participation Group, Instituto de Astrofísica de Canarias, Johns Hopkins University, Kavli Institute for the Physics and Mathematics of the Universe (IPMU)/University of Tokyo, the Korean Participation Group, Lawrence Berkeley National Laboratory, Leibniz Institut für Astrophysik Potsdam (AIP), Max-Planck-Institut für Astronomie (MPIA Heidelberg), Max- Planck-Institut für Astrophysik (MPA Garching), Max-Planck- Institut für Extraterrestrische Physik (MPE), National Astro- nomical Observatories of China, New Mexico State University, New York University, University of Notre Dame, Observatário Nacional/MCTI, Ohio State University, Pennsylvania State University, Shanghai Astronomical Observatory, United King- dom Participation Group, Universidad Nacional Autónoma de México, University of Arizona, University of Colorado Boulder, University of Oxford, University of Portsmouth, University of Utah, University of Virginia, University of Washington, University of Wisconsin, Vanderbilt University, and Yale University.

The Hyper Suprime-Cam (HSC) collaboration includes the astronomical communities of Japan and Taiwan as well as Princeton University. The HSC instrumentation and software were developed by the National Astronomical Observatory of Japan (NAOJ), the Kavli Institute for the Physics and Mathematics of the Universe (Kavli IPMU), the University of Tokyo, the High Energy Accelerator Research Organization (KEK), the Academia Sinica Institute for Astronomy and Astrophysics in Taiwan (ASIAA), and Princeton University.
Funding was contributed by the FIRST program from the Japanese Cabinet Office, the Ministry of Education, Culture, Sports, Science and Technology (MEXT), the Japan Society for the Promotion of Science (JSPS), Japan Science and Technol- ogy Agency (JST), the Toray Science Foundation, NAOJ, Kavli IPMU, KEK, ASIAA, and Princeton University. This paper makes use of software developed for the Large Synoptic Survey Telescope (LSST). We thank the LSST Project for making their code available as free software at http://dm.lsst. org. This paper is based [in part] on data collected at the Subaru Telescope and retrieved from the HSC data archive system, which is operated by Subaru Telescope and Astronomy Data Center (ADC) at National Astronomical Observatory of Japan. Data analysis was in part carried out with the cooperation of Center for Computational Astrophysics (CfCA), National Astronomical Observatory of Japan.

\facilities{MMT Hectospec, Subaru Hyper Suprime Cam}

\bibliographystyle{aasjournal}
\bibliography{ms}

\end{document}